%% file: main.tex
\documentclass[letterpaper,twocolumn,10pt]{article}
\usepackage{usenix-2020-09}

\usepackage{tikz}
\usepackage{amsmath}



\usepackage{filecontents}

\makeatother
\usepackage{hyperref}
\usepackage{multirow}
\usepackage[utf8]{inputenc}
\usepackage[ruled, lined, longend, linesnumbered]{algorithm2e}

\usepackage{array}
\newcolumntype{L}[1]{>{\raggedright\let\newline\\\arraybackslash\hspace{0pt}}m{#1}}
\newcolumntype{C}[1]{>{\centering\let\newline\\\arraybackslash\hspace{0pt}}m{#1}}
\newcolumntype{R}[1]{>{\raggedleft\let\newline\\\arraybackslash\hspace{0pt}}m{#1}}

\input{header}

\newcommand{\mwx}[1]{{\color{red}{#1}}}
\newcommand{\revision}[1]{{#1}}

\usepackage{authblk}

\newcommand\blfootnote[1]{%
	\begingroup
	\renewcommand\thefootnote{}\footnote{#1}%
	\addtocounter{footnote}{-1}%
	\endgroup
}

\begin{document}
	
	\title{Video Analytics with Zero-streaming Cameras}

\author[1,2 *]{Mengwei Xu}
\author[3 *]{Tiantu Xu}
\author[4]{Yunxin Liu}
\author[5]{Felix Xiaozhu Lin}
\affil[1]{\textit{Peking University}}
\affil[2]{\textit{Beijing University of Posts and Telecommunications}}
\affil[3]{\textit{Purdue ECE}}
\affil[4]{\textit{Institute for AI Industry Research (AIR), Tsinghua University}}
\affil[5]{\textit{University of Virginia}}

\maketitle


	\blfootnote{*Mengwei Xu and Tiantu Xu contributed equally to the paper.}
	\blfootnote{*Work done during Mengwei Xu's visit to Purdue University.}

	\input{abstract}
	\thispagestyle{plain}
	\pagestyle{plain}

\input{todo}\input{intro}   			  
	\input{bkgnd}         
	\input{overview}			  
	\input{landmark}
\input{op}

	\input{design}
	\input{impl}			      
	\input{eval}
	\input{related}
	\input{conclusions}      
	
	\section*{Acknowledgments}
	Mengwei Xu was supported by National Key R\&D Program of China under grant number 2020YFB1805500, the Fundamental Research Funds for the Central Universities, and National Natural Science Foundation of China under grant numbers 62032003, 61922017, and 61921003.
	Tiantu Xu and Felix Xiaozhu Lin were supported in part by NSF awards \#1846102 and \#1919197.
	We thank our shepherd, Michael Kozuch, and the anonymous ATC reviewers for their useful suggestions.

	\bibliographystyle{plain}
	\interlinepenalty=10000 
	\bibliography{bib/abr-short,bib/xzl,bib/misc,bib/cv,bib/book,mwx.bib}
	
	
\end{document}

%% file: header.tex
\usepackage{wrapfig}
\usepackage{comment}
\usepackage{colortbl}
\usepackage{tabularx}

\usepackage{color,soul}
\usepackage{graphics}
\usepackage{graphicx}
\usepackage{lipsum}

\usepackage{float} 

\usepackage{CJKutf8} 

\usepackage{tikz}
\usetikzlibrary{decorations.pathmorphing, patterns,shapes,snakes} 

\usepackage[inline]{enumitem}	
\usepackage{listings} 
\usepackage{lipsum}

\usepackage{capt-of}	

\usepackage{ulem}  

\ifdefined\HCode
\usepackage[compatibility=false]{caption}

\else
\usepackage{subcaption} 
\fi

\usepackage{mathrsfs}	
\usepackage{amsfonts}

\usepackage[export]{adjustbox} 

\definecolor{myForestGreen}{RGB}{34, 139, 34}
\definecolor{airforceblue}{rgb}{0.36, 0.54, 0.66}
\newcommand*\circled[1]{\tikz[baseline=(char.base)]{
		\node[shape=circle,fill=black,draw,text=white,inner sep=1pt] (char) {#1};}}

\definecolor{refkey}{RGB}{34,139,34}
\definecolor{labelkey}{rgb}{0,0,1}

\renewcommand{\paragraph}[1]{\vskip 3pt\noindent\textbf{#1 }}	 

  {\begin{list}{$\bullet$}%
     {\setlength{\parsep}{0pt}%
      \setlength{\topsep}{0pt}%
      \setlength{\itemsep}{2pt}}}%
  {\end{list}}
\newcommand\note[1]{\sethlcolor{yellow} \hl{#1}} 
\newcommand\Note[1]{\sethlcolor{green} \hl{#1}} 

\newcommand\notettx[1]{\sethlcolor{orange} \hl{#1}} 

\newcommand{\textbfit}[1]{\textbf{\textit{#1}}}
		
\newcommand\sect[1]{$\S$\ref{sec:#1}}	
		
\newenvironment{myitemize}%
  {\begin{itemize}
	[leftmargin=0cm,
		itemindent=.3cm,
		labelwidth=\itemindent,
		labelsep=0pt,
		parsep=3pt,
		topsep=2pt,
		itemsep=1pt,
		align=left]
  }%
  {\end{itemize}}    

\newenvironment{myenumerate}%
  {\begin{enumerate}
	[leftmargin=0cm,itemindent=.5cm,labelwidth=\itemindent,
		labelsep=0pt,
		parsep=1pt,
		topsep=1pt,
		itemsep=3pt,
		align=left]
  }%
  {\end{enumerate}}    
  
\newenvironment{enumerateinline}%
  {\begin{enumerate*}
	[label=(\arabic*)]
  }%
  {\end{enumerate*}}      

\newenvironment{myenumerateinline}%
  {\begin{enumerate*}
	[label=(\arabic*)]
  }%
  {\end{enumerate*}}

\newcommand{\code}[1]{\texttt{\small{#1}}}	

\input{terms}

\makeatletter
\def\@copyrightspace{\relax}
\makeatother

%% file: terms.tex
\newcommand{\sys}{DIVA}

\newcommand{\jacksonh}{JacksonH}
\newcommand{\jacksont}{JacksonT}
\newcommand{\banff}{Banff}
\newcommand{\mierlo}{Mierlo}
\newcommand{\miami}{Miami}
\newcommand{\ashland}{Ashland}
\newcommand{\shibuya}{Shibuya}

\newcommand{\chaweng}{Chaweng}
\newcommand{\lausanne}{Lausanne}
\newcommand{\venice}{Venice}
\newcommand{\tucson}{Tucson}
\newcommand{\oxford}{Oxford}
\newcommand{\whitebay}{Whitebay}

\newcommand{\coralreef}{CoralReef}
\newcommand{\boathouse}{BoatHouse}

\newcommand{\eagle}{Eagle}

\newcommand{\countmax}{max Count}

\newcommand{\OpF}{E\textsubscript{CHEAP}}
\newcommand{\OpS}{E\textsubscript{EXP}}

\newcommand{\cloudonly}{\code{CloudOnly}}
\newcommand{\noscope}{\code{OptOp}}
\newcommand{\focus}{\code{PreIndexAll}}


%% file: abstract.tex
\subsection*{Abstract}
Low-cost cameras enable powerful analytics. 
An unexploited opportunity is that most captured videos remain ``cold'' without being queried. For efficiency, we advocate for these cameras to be \textit{zero streaming}: capturing videos to local storage and communicating with the cloud only when analytics is requested. 

How to query zero-streaming cameras efficiently?
Our response is a camera/cloud runtime system called \sys{}. 
It addresses two key challenges: 
to best use limited camera resource during video capture; 
to rapidly explore massive videos during query execution. 
\sys{} contributes two unconventional techniques. 
(1) When capturing videos, a camera builds sparse yet accurate landmark frames, from which it learns reliable knowledge for accelerating future queries. 
(2) When executing a query, 
a camera processes frames in multiple passes with increasingly more expensive operators. 
As such, \sys{} presents and keeps refining inexact query results throughout the query's execution. 
On diverse queries over 15 videos lasting 720 hours in total, 
\sys{} runs at more than 100$\times$ video realtime and outperforms competitive alternative designs. 
To our knowledge, \sys{} is the first system for querying large videos stored on low-cost remote cameras. 

%% file: intro.tex
\section{Introduction}
\label{sec:intro}

Cameras are pervasive: 
a survey of 61 organizations shows that from
2015 to 2018 their average number of cameras has increased by almost 70\%, from 2,900 to 4,900~\cite{camera-survey}.
Insights of videos can be extracted by queries such as 
``get the daily peak pedestrian count in the past week''~\cite{lipton2015video,zhu2017online,shi2018geometry,beymer1997real}. 
Four recent trends motivate our work.

\noindent 
\textit{(1) Low-cost, wireless cameras grow fast}
\hspace{1mm}
As key complements to high-end cameras, 
low-cost cameras (<\$40) are increasingly pervasive~\cite{wyze-camv2,yi-cam,zosi-cam}. 
These cameras often have limited compute resources yet spacious storage. 
Being wireless, these cameras are meant to be installed by individuals or small businesses with ease just as other wireless sensors. 


\noindent \textit{(2) Most videos are cold} \hspace{1mm}
Users deploy cameras to knowingly capture excessive videos, expecting that most videos will never be queried~\cite{wdblog}.
This is because interesting events are often unforeseeable, e.g., car accidents; 
the need for examining such events emerges well after the fact.
\sect{bkgnd:cold} presents a 6-month study of real-world camera deployment, where only <0.005\% of captured videos are eventually queried. 

\input{tab-scenario}


\noindent \textit{(3) Transmitting cold videos wastes wireless bandwidth}
\hspace{1mm}
Cold videos should not compete with human users for network bandwidth. 
If streaming video in real-time, a single camera generates traffic at 0.2 MB/s--0.4 MB/s  (720P@1--30 FPS);
with multiple cameras on one network, their always-on streams easily consume most, if not all, bandwidth of consumer WiFi, which is 0.2 MB/s--3 MB/s (median: 0.99) in a recent global survey~\cite{wifi-state} and less than 1.5 MB/s in an academic study~\cite{mpdash}. 
%
A dedicated network for cameras is expensive, as the network monetary cost will exceed the camera cost in several months~\cite{Comcast-dataplan}. 

\noindent \textit{(4) Camera storage can retain videos long enough}
\hspace{1mm}
A cheap camera can already store videos for weeks or months. 
Such retention periods already satisfy many video  scenarios~\cite{surveillance-policy,surveillance-policy2}.
In fact, legal regulations often \textit{prevent}  retention longer than a few months, mandating video deletion for privacy~\cite{retention-case-law,eu-video-guideline}.
Existing measures can assure data security of on-camera videos. 
\sect{bkgnd:case} will provide evidence in detail.

\paragraph{Zero streaming \& its use cases}
How to analyze cold videos produced by numerous low-cost cameras?
We advocate for a system model dubbed ``zero streaming''. 
(1) Cameras continuously capture videos to their local storage without uploading any. 
(2) Only in response to a retrospective query, 
the cloud reaches out to the queried camera and coordinates with it to process
the queried video. 
(3) While the video is being processed, the system presents users with inexact yet useful results; it continuously refines the results until query completion~\cite{ola}. 
In this way, a user may \textit{explore} videos through interactive queries, e.g., aborting an ongoing query based on inexact results and issuing a new query with revised parameters~\cite{eva,eureka}.
Zero streaming has rich use cases, for example:

\begin{myitemize}

\item 
To trace the cause of recent frequent congestion on a highway, 
a city planner queries cameras on nearby local roads, requesting car counts seen on these local roads.

\item 
To understand how recent visitors impact bobcat activities, a ranger queries all the park's cameras, requesting time ranges where the cameras capture bobcats.

\end{myitemize}

\paragraph{Advantages}
Zero streaming suits resource-frugal cameras in large deployment. 
When capturing videos, cameras require no network or external compute resources.
Only to process a query, the cameras require networks such as long-range wireless~\cite{lora} and cloud resources such as GPU. 
Zero streaming adds a new point to the design space of video analytics shown in Figure~\ref{fig:scenario}. 
It facilitates retrospective, exploratory analytics, a key complement to real-time event detection and low-delay video retrieval~\cite{reducto, videostorm,focus,chameleon}. 
The latter demands higher compute or network resources per camera and hence suits fewer cameras around hot locations such as building entrances.

\paragraph{\sys{}}
To support querying zero-streaming cameras, 
we present a camera/cloud runtime called \sys{}. 
As shown in Figure~\ref{fig:overview},
a camera captures video to local storage; 
it deletes videos after their maximum retention period. 
In response to a query, the camera works in conjunction with the cloud:
the camera runs operators, implemented as lightweight neural nets (NNs), to \textit{rank} or \textit{filter} frames; 
the cloud runs full-fledged object detection to validate results uploaded from the camera.  
\sys{} thus does not sacrifice query accuracy, ensuring it as high as that of object detection by the cloud. 

\input{fig-overview}

The major challenges to \sys{} are two. 
(1) During video capture: how should cameras best use limited resources for future queries? 
(2) To execute a query: how should the cloud and the camera orchestrate to deliver useful results rapidly?
%
%
Existing techniques are inadequate.
Recent systems pre-process (``index'') video frames as capturing them~\cite{focus}
and answer queries based on indexes only. 
Yet, 
as we will show in \sect{eval}, 
low-cost cameras can hardly build quality indexes in real-time.
Many systems process video frames in a streaming fashion~\cite{vigil,lavea,filterforward,glimpse,cloudseg}, 
which however miss key opportunities in retrospective queries. 
\noindent To this end, \sys{} has two unconventional designs. 


\begin{myitemize}
\item 
\textbf{During video capture: building sparse but sure landmarks 
to distill long-term knowledge}
(Figure~\ref{fig:overview}(a)) 
To optimize future queries, 
our key insight is that \textit{accurate} knowledge on a \textit{sparse} sample of frames is much more useful than \textit{inaccurate} knowledge on \textit{all} frames.
This is opposite to existing designs that detect objects with low accuracy on all/most frames as capturing them~\cite{focus,filterforward}. 
On a small sample of captured video frames dubbed \textit{landmarks}, 
the camera runs generic, expensive object detection, e.g., YOLOv3~\cite{yolov3}. 
Constrained by camera hardware, landmarks are sparse in time, e.g., 1 in every 30 seconds; 
yet, with high-accuracy object labels, they provide reliable spatial distributions of various objects over long  videos.
High accuracy is crucial, as we will validate through evaluation ($\S$\ref{sec:eval:lm}).
\sys{} optimizes queries with landmarks: 
it prioritizes processing of frame regions with object skewness learned from landmarks;
it bootstraps operators with landmarks as training samples.
Landmarks only capture a small fraction of object instances; those
uncaptured do not affect correctness/accuracy (\S\ref{sec:lm}).


\item 
\textbf{To execute queries: multipass processing with online operator upgrade}
(Figure~\ref{fig:overview}(b)) 
To process large videos, our key insight is to refine query results in multiple passes, each pass with a more expensive/accurate operator. 
Unlike prior systems processing all frames in one pass and delivering results in one shot~\cite{noscope,filterforward,blazeit}, 
multipass processing produces useful results during query execution, enabling users to explore videos effectively.
To do so, 
\sys{}'s cloud trains operators with a wide spectrum of accuracies/costs. 
Throughout query execution, the cloud keeps pushing new operators to the camera, picking the next operator based on query progress, network conditions, 
and operator accuracy. 
The early operators quickly \textit{explore} the frames for inexact answers 
while later operators slowly \textit{exploit} for more exact answers. 

\end{myitemize}




On 720-hour videos in total from 15 different scenes,  \sys{} runs queries at more than 100$\times$ video realtime on average, with typical wireless conditions and low-cost hardware.
\sys{} returns results quickly: 
compared to executing a query to completion, 
\sys{} takes one order of magnitude shorter time to return half of the result frames. 
Compared to competitive alternatives, \sys{} speeds up queries by at least 4$\times$.

\paragraph{Contributions}
We have made the following contributions. 
\begin{myitemize}

\item Zero streaming,  a new model for low-cost cameras to operate on frugal networks while answering video queries.  

\item Two novel techniques for querying zero-streaming cameras:
optimizing queries with accurate knowledge from sparse frames; 
processing frames in multiple passes with operators continuously picked during a query. 

\item \sys{}, a concrete implementation that runs queries at more than 100$\times$ realtime with uncompromised query accuracy.
To our knowledge, \sys{} is the first system designed for querying large videos stored on low-cost remote cameras. 
\end{myitemize}


\paragraph{Ethical considerations}
In this study: 
all visual data used is from the public domain;
no information traceable to human individuals is collected or analyzed. 

%% file: tab-scenario.tex

\begin{figure}[t!]
\centering
\includegraphics[width=0.45\textwidth{}]{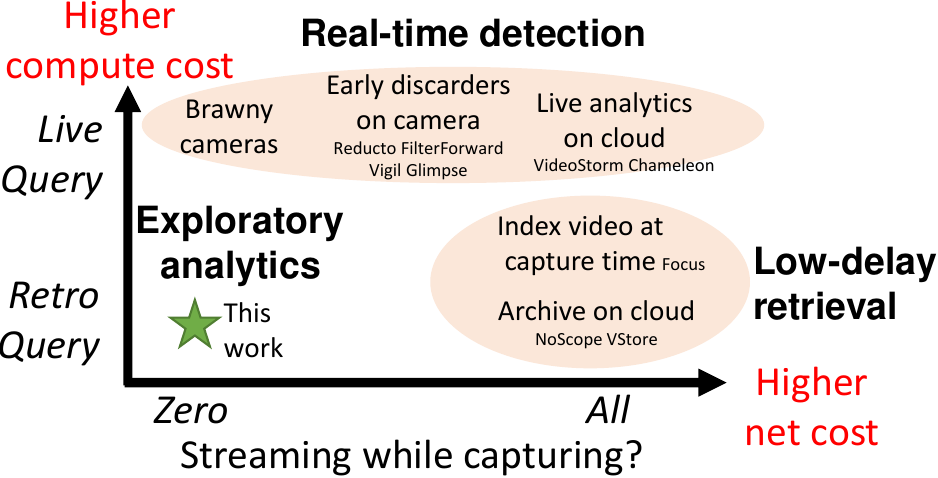} 
\caption{\textbf{The design space} of video analytics systems, showing this work and prior systems.
}
\label{fig:scenario}
\vspace{-10pt}		
\end{figure}

%% file: fig-overview.tex

\begin{figure}[t]
	\centering
	\includegraphics[width=0.4\textwidth]{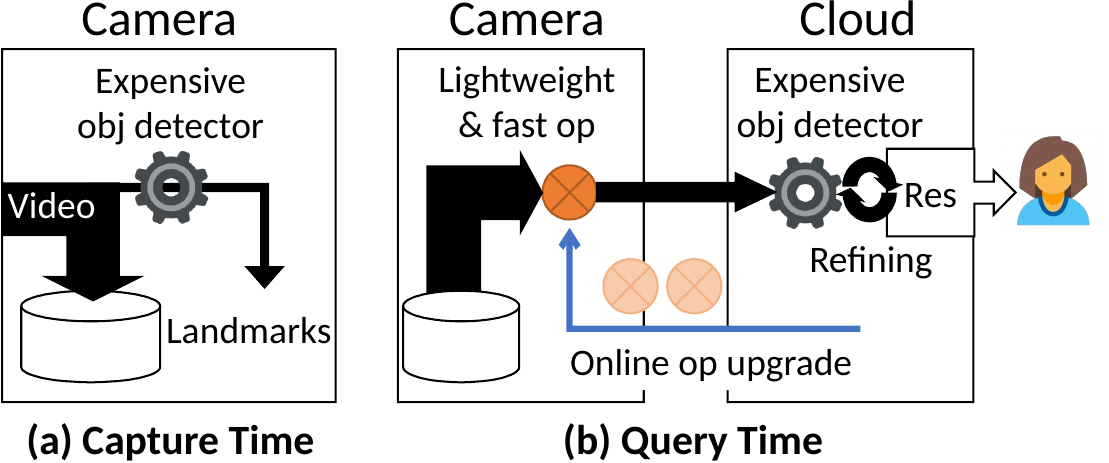}
	\vspace{-5pt}
	\caption{\textbf{Overview of \sys{}}}
	\label{fig:overview}
\vspace{-5pt}		
\end{figure}


%% file: bkgnd.tex

\section{Motivations}
\label{sec:bkgnd}



\subsection{Cold videos are already pervasive}\label{sec:bkgnd:cold}

\paragraph{Case study: Cold videos in real-world deployment}
We conduct an IRB-approved study examining existing camera deployment on PKU campus.
Spanning 1 mi$^2$, the campus hosts tens of thousands of employees 
and operates more than 1,000 cameras. 
All captured videos are stored for a few months for retrospective queries before deletion.
The camera deployment supports AI-based queries, e.g., object detection,  \textit{not} traceable to unique persons, and reviews by human analysts.
We analyzed system logs spanning six continuous months:
in over 3,000,000 hours of videos (5.4 PB) have been captured, only {<0.005\%} video data from {<2\%} cameras are queried.

\paragraph{Why are most videos cold?}
(1) Interesting video events are both unpredictable (thus the need for capturing excessive videos) and sparse (thus low chances for footage being queried). 
For example, severe traffic breakdown contributes to less than 5\% of the time per day~\cite{trafficpercentage}; 
Foreign intelligence surveillance court only reviewed a tiny fraction of video for terrorism events~\cite{terrorismpercentage}.
(2) Analyzing videos is expensive:
it still requires a GPU of a few thousand dollars for high-accuracy object detection over a video stream~\cite{noscope}. 
(3) In years to come, cheap cameras will produce more videos. 

\subsection{Target queries and their execution}
\label{sec:back:evidence}




We target ad-hoc queries~\cite{focus,noscope,vstore,vigil}.
The query parameters, including object classes, video timespans, and expected accuracies, are specified at query time rather than video capture time. 
Such queries are known for flexibility.

\paragraph{High-accuracy object detection is essential}
Object detection is the core of ad-hoc queries~\cite{blazeit}.
Minor accuracy loss in object detection may result in substantial loss in query performance, as we will demonstrate in \sect{eval}. 
While NNs significantly advance object detection, 
new models with higher accuracy demand much more compute. 
For instance, compared to YOLOv3 (2018)~\cite{yolov3},
CornerNet (2019)~\cite{cornernet} improves Average Precision by 28\% while being 5$\times$ more expensive.

\paragraph{Low-cost cameras cannot answer queries without cloud}
Cameras in real-world deployment are reported to be resource-constrained~\cite{reducto}.
Low-cost cameras (<\$40) have wimpy cores, e.g., Cortex-A9 cores for YI Home Camera~\cite{yi-cam} and MIPS32 cores for WyzeCam~\cite{wyze-camv2}; 
their DRAM is no more than a few GBs~\cite{wyzecam, cheapcam}. 
In recent benchmarks, they run state-of-the-art object detection at 0.1 FPS~\cite{nnbenchmarks,yolov2-rpi}, incapable of keeping up with video capture at 1--30 FPS~\cite{focus,noscope}. 
NN accelerators
still cannot run high-accuracy object detection fast enough at low enough monetary cost, e.g., Intel's Movidius (\$70) runs YOLOv3 at no faster than 0.5 FPS. 
In the foreseeable future, we expect that the resource gap between high-accuracy object detection and low-cost camera continues to exist. 







\subsection{A case for zero streaming}\label{sec:bkgnd:case}

\paragraph{Streaming cold videos wastes bandwidth}
As discussed in \sect{intro}, cameras are cheap while wireless spectrum is precious. 
Deploying streaming cameras on a shared network incurs poor experience~\cite{complain-camera-1,complain-camera-2} and draws researcher attention~\cite{vigil,filterforward}. 
Dedicated networks are costly~\cite{Comcast-dataplan} and thus only suit a small number of cameras in critical locations. 
While wireless bandwidth grows, consumer demand grows even faster, e.g., 20$\times$ for VR/AR and 10$\times$ for gaming~\cite{cisco-white-paper}.
Cold video traffic should not contend with consumers for network bandwidth.

\paragraph{Streaming optimizations cannot offset the waste} 
One may reduce FPS or resolution of streamed videos. 
Even if users tolerate the resultant lower query accuracy, 
the saved bandwidth is incomparable to the waste on overwhelmingly streamed cold videos, as we will experimentally show (\S\ref{sec:eval}).
On-camera ``early filters'' ~\cite{filterforward,reducto,glimpse} are still suboptimal when querying massive \textit{cold} videos.
(1) Without knowing query objects/parameters at video capture time, 
a camera may run a generic filter, e.g., discarding no-motion frames; 
it still streams substantial survival frames (e.g., consider a street-view camera). 
As stated above, most of these frames will remain cold and hence wasted. 
(2) The camera may run a large set of specific filters covering all possible query objects/parameters.
Even if possible, this incurs a much higher compute cost to camera. 

\paragraph{Edge processing does not justify streaming}
Cameras may stream to edge servers. 
Yet, streaming hundreds if not thousands of \textit{always-on}, \textit{cold} video streams, even if possible on certain wireless infrastructures, still wastes precious wireless spectrum at the edge~\cite{spectrum-crunch}.
Furthermore, deploying and managing video edge servers can be challenging and costly in many scenarios, 
such as construction sites and remote farms. 

\input{tab-storage}

\paragraph{Cameras can retain videos long enough}
Table~\ref{tab:storage} shows the price of $\mu$SD cards has been dropped by 2.6$\times$--5.4$\times$ in the past few years.
Cameras can retain videos for several weeks and for several months soon.
Such a retention period is already adequate for most retrospective query scenarios, where videos are retained from a few weeks to a few months based on best practice and legal regulations~\cite{surveillance-policy,surveillance-policy2,retention-case-law,eu-video-guideline}.
For privacy, 
many regulations \textit{prohibit} video retention longer than a few months and mandate deletion afterwards~\cite{retention-case-law,eu-video-guideline}.

\paragraph{Our model \& design scope} 
To harness cold videos, we advocate for zero streaming. 
We focus on cold videos being queried for the first time and querying individual cameras.
We intend our design to form the basis of future enhancement and extension, e.g., resource scheduling for multiple queries/users/tenants~\cite{filterforward}, caching for repetitive queries~\cite{xu2018deepcache}, exploiting past queries for refinement~\cite{queryrefinement}, and exploiting cross-camera topology~\cite{rexcam}. 
We address limited compute resource on cameras~\cite{cheapcam} and limited network bandwidth~\cite{mpdash}. 
We do not consider the cloud as a limiting factor, 
assuming it runs fast enough to process frames uploaded from cameras.

%% file: tab-storage.tex


\begin{table}[H]
\small
\begin{tabular}{rrrll}
\hline
Size & Yr.2017 & Yr.2020 & 720p@30FPS & 720p@1FPS \\ \hline
128GB & \$45 & \$17 & $\sim$11 days &  $\sim$3 weeks \\
256GB & \$150 & \$28 & $\sim$3 weeks & $\sim$ 6 weeks \\ \hline
\end{tabular}
\caption{Cheap $\mu$SD cards on cameras retain long videos for humans to review~\cite{fortinet-white-paper} or for machines to analyze~\cite{focus}.
}
\label{tab:storage}
	\vspace{-10pt}
\end{table}

%% file: overview.tex

\section{The \sys{} Overview}
\label{sec:overview}

\input{query}

\input{fig-design-execution}

\paragraph{Query execution}
Upon receiving a query, the cloud retrieves all landmarks in queried video as low-resolution thumbnails, e.g., 100$\times$100, with object labels and bounding boxes
(Figure~\ref{fig:design-execution} \circled{1}).
The cloud uses landmarks:
\textbf{\textit{(1) to estimate object spatial distribution}}, e.g., ``90\% queried objects appear in a 100$\times$100 region on the top-right'', which is crucial to query optimization (\S\ref{sec:lm});
\textbf{\textit{(2) as the initial training samples}}
for bootstrapping a family of camera operators (\circled{2}).  
The camera filters/ranks frames and uploads the ranked or surviving frames (\circled{3}). 
The cloud processes the uploaded frames and emits results, e.g., positive frames.
It trains operators for higher accuracy (\circled{4}).
Observing resource conditions and positive ratios in uploaded frames,
the cloud upgrades the operator on camera (\circled{5}).  
With the upgraded operator, the camera continues to process remaining frames (\circled{6}).
Step (\circled{4})--(\circled{6}) repeat until query abort or completion. 
Throughout the query, the cloud keeps refining the results presented to the user (\circled{7}).

\paragraph{Notable designs}
(1)
The camera processes frames in multiple passes, one operator in each pass. 
(2) 
The camera processes and uploads frames asynchronously.
For instance, 
when the camera finishes ranking 100 out of total 1,000 frames, it may have uploaded the top 50 of the 100 ranked frames. 
This is opposed to common ranking which holds off frame upload until all the frames are ranked~\cite{rank-aware-opt,similarity-join,koudas2000high}. 
(3) 
The processing/upload asynchrony facilitates video exploration: 
it amortizes query delay over many installments of results;
it pipelines query execution with user thinking~\cite{eureka}.
Table~\ref{tab:query} summarizes a user's view of query results and the performance metrics.
While such online query processing has been known~\cite{olamr,online-mapreduce},
we are the first applying it to visual data. 

\paragraph{Limitations} 
\sys{} is not designed for several cases and may underperform:
querying very short video ranges, e.g., minutes, for which simply uploading all queried frames may suffice without operators; 
querying non-stationary cameras for which landmarks may not yield accurate object distribution.
\sys{} is vulnerable to loss of video data in case of camera storage failure. Users can mitigate such a risk via cross-camera data backup (RAID-like techniques) on the same local area network or by increasing camera deployment density.

%% file: query.tex


\input{tab-query}

\paragraph{Query types}
Concerning a specific camera, 
an ad-hoc query ($\mathcal{T, C}$) covers a video timespan $\mathcal{T}$, typically hours or days, and 
an object class $\mathcal{C}$ as detectable by modern NNs, e.g., any of the 80 classes of YOLOv3~\cite{yolov3}.
As summarized in Table~\ref{tab:query}, 
\sys{} supports three query types:
\textbf{\textit{Retrieval}}, 
e.g., ``retrieve all images that contain buses from yesterday'';
\textbf{\textit{Tagging}}, 
e.g., ``return all time ranges when any deer shows up in the past week'', 
in which the time ranges are returned as metadata but not images;
\textbf{\textit{Counting}},
e.g., ``return the maximum number of cars that ever appear in any frame today''. 

\paragraph{System components}
\sys{} spans a camera and the cloud.
Between them, the network connection is only provisioned at query time.
To execute a query,
a camera runs lightweight NNs, or operators, to \textit{filter} or \textit{rank} the queried frames for upload. 
On the uploaded frames, 
the cloud runs generic, high-accuracy object detection and  materializes query results. 
Table~\ref{tab:query} summarizes executions for different queries:

\noindent
$\bullet$
The camera executes \textbfit{rankers} for \textit{Retrieval} and \textit{\countmax} queries.
A ranker scores frames; a higher score suggests that a frame is more likely to contain \textit{any} object of interest (for Retrieval) or \textit{a large count} of such objects (for \countmax{}).

\noindent
$\bullet$
The camera executes \textbfit{filters} for \textit{Tagging} queries.
A filter scores frames;
it resolves any frame scored below/above two pre-defined 
thresholds as negative/positive, 
and deems other frames as unresolved. 
For each resolved frame, the camera uploads a positive/negative tag; 
the camera either uploads unresolved frames for the cloud to decide or defer them to more accurate filters on camera in subsequent passes.

%% file: tab-query.tex

\begin{table*}[t!]
\centering
\includegraphics[width=0.98\textwidth{}]{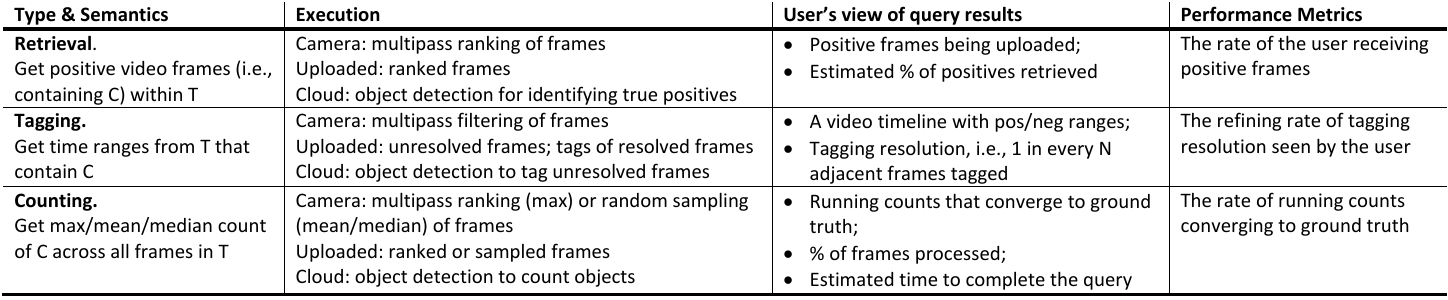} 
\vspace{-5pt}
\caption{\textbf{A summary of supported queries.}
$\mathcal{T}$ is the queried video timespan; 
$\mathcal{C}$ is the queried object class
}
\label{tab:query}
\vspace{-5pt}		
\end{table*}

%% file: fig-design-execution.tex
\begin{figure}[t]
	\centering
	\includegraphics[width=0.4\textwidth]{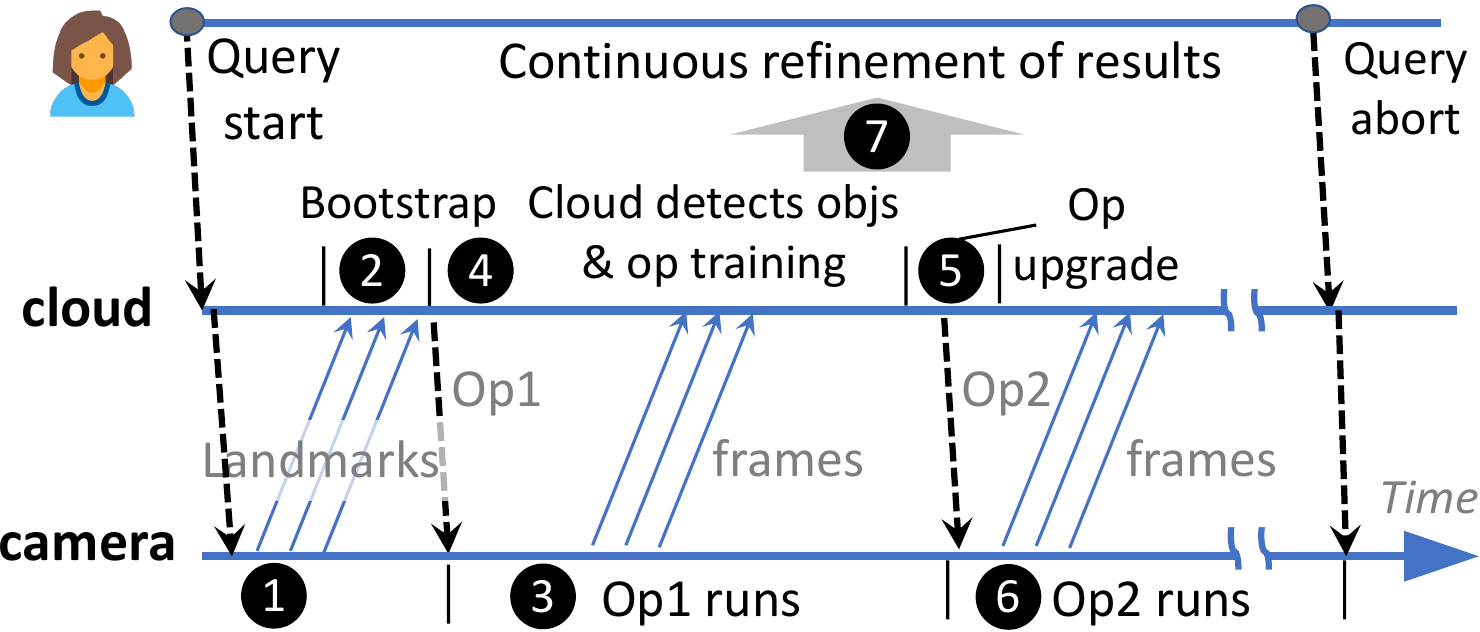}
	\caption{\textbf{The workflow of a query's execution.}}
	\vspace{-5pt}
	\label{fig:design-execution}
\end{figure}

%% file: landmark.tex


\section{Landmark Design}
\label{sec:lm}



\input{fig-hm-combine}
\input{fig-hm-time}

Surveillance cameras have a unique opportunity: 
to learn \textit{object class distribution} from weeks of videos.  
We focus on \textbf{spatial skews}:
objects of a given class are likely to concentrate on certain small regions on video frames. 
In examples of Figure~\ref{fig:hm-combine}(a)-(b),
most cars appear near a stop sign;
most persons appear in a shop's aisle.
Such long-term skews are rarely tapped in prior computer vision work, which mostly focused on minute-long videos~\cite{videoedge,rexcam,saligrama12cvpr,zhu18cvpr,shen17cvpr}.
Compared to recent work that improved classifier performance by cropping video frames~\cite{filterforward}, \sys{} takes a step further by automatically learning spatial skews from sparse frames with resource efficiency.

The design is backed up by three key observations. 
(1)
One object class may exhibit different skews in different videos (Figure~\ref{fig:hm-combine}(a)-(c));
different classes may exhibit different skews in the same video (Figure~\ref{fig:hm-combine}(c)). 
(2)
The skews are pervasive: 
surveillance cameras cover long time spans and a wide field of view, where objects are small;
in the view, objects are subject to social constraints, e.g., buses stop at traffic lights,
or physical constraints, e.g., humans appear on the floor. 
(3)
The skews can be learned through \textit{sparse} frame samples, as exemplified by Figure~\ref{fig:hm-time}. 





To exploit such an opportunity, \sys{} makes the following design choices.
\textbf{High-accuracy object detection:}
at capture time, the camera runs an object detector with the highest accuracy as allowed by the camera's hardware, mostly memory capacity. 
This is because camera operators crucially depend on the correctness of landmarks, i.e., the object labels and bounding boxes.
We will validate this experimentally (\sect{eval:lm}). 
\textbf{Sparse sampling at regular intervals:}
to accommodate slow object detection on cameras, 
the camera creates landmarks at long intervals, 
e.g., 1 in every 30 seconds in our prototype (\S\ref{sec:eval}). 
Sparse sampling is proven valid for estimating statistics of low-frequency signals~\cite{sparsesampling}, e.g., object occurrence in videos in our case. 
We will validate this (\sect{eval:lm});
without assuming a priori of object distribution, regular sampling ensures unbiased estimation of the distribution~\cite{regularsampling}.
Given a priori, cameras may sample at corresponding random intervals for unbiased estimation. 

\paragraph{Key idea: exploiting spatial skews for performance}
The cloud learns the object class distribution from landmarks of the queried video timespan. 
It generates a heatmap for spatial distribution (Figure~\ref{fig:hm-combine}).
Based on the heatmap, 
the cloud produces camera operators consuming frame regions of different locations and sizes.
Take Figure~\ref{fig:hm-combine}(a) as an example:
a filter may consume bottom halves of all frames and accordingly filter frames with no cars; 
for Figure~\ref{fig:hm-combine}(b),  
a ranker may consume a smaller bounding box where 80\% persons appear and rank frames based on their likelihood of containing more persons. 
Figure~\ref{fig:op-spd-acc} shows that, 
by zooming into smaller regions,
operators run faster and deliver higher accuracy.
By varying input region locations/sizes, 
\sys{} produces a set of operators with diverse costs/accuracies. 
By controlling the execution order of operators, \sys{} processes ``popular'' frame regions prior to ``unpopular'' regions. 
\sys{} never omits any region when it executes a query to completion to guarantee correctness.

\paragraph{What happens to instances uncaptured by landmarks?}
Sparse by design, landmarks are not meant to capture all object instances;
instead, they are used as inexact estimators and initial training samples.
Reducing landmarks will degrade query speed, as we will experimentally quantify in $\S$\ref{sec:eval:lm}.  
Doing so, however, does not affect query correctness or accuracy: 
the instances uncaptured by landmarks will be eventually processed by \sys{} as a query goes on.


\input{fig-op-spd-acc}

%% file: fig-hm-combine.tex
\begin{figure}[t]
	\centering
	\includegraphics[width=0.48\textwidth]{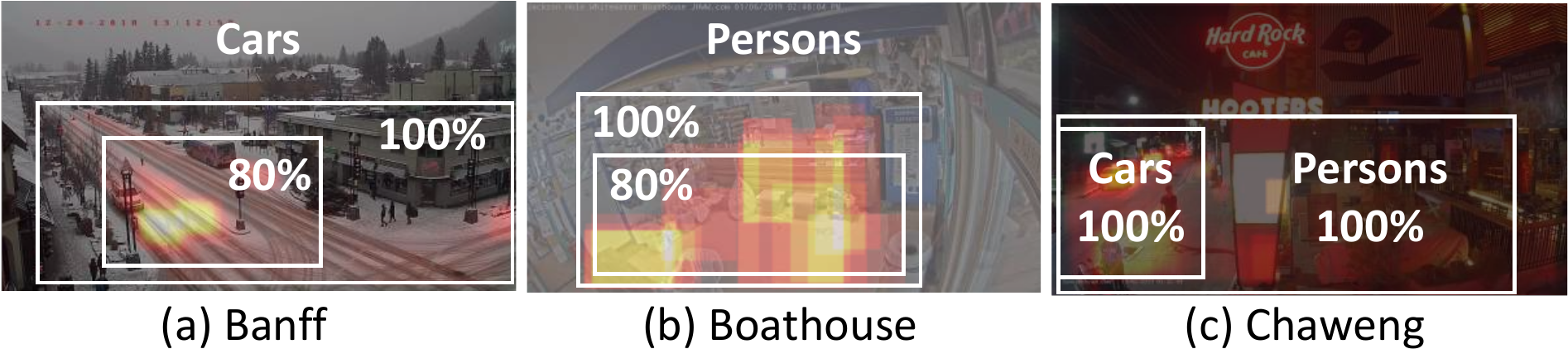}
	\caption{
	\textbf{Class spatial skews in videos}.
	In (a) \banff{}: 80\% and 100\% of cars appear in regions that are only 19\% and 57\% of the whole frame, respectively. 
	}
	\label{fig:hm-combine}
		\vspace{-10pt}
\end{figure}

%% file: fig-hm-time.tex
\begin{figure}[t]
	\centering
	\includegraphics[width=0.48\textwidth]{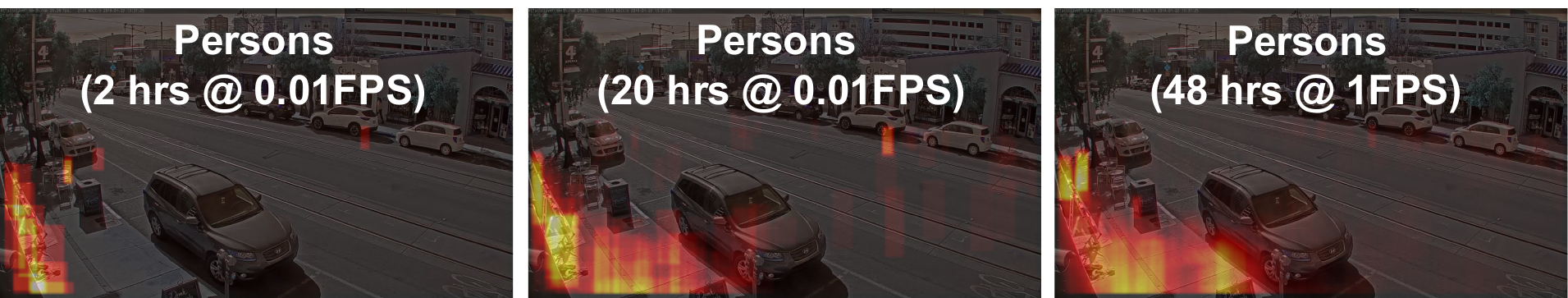}
	\vspace{-5pt}
	\caption{
	\textbf{Class spatial distribution can be estimated from 
	sparse frames sampled over long video footage.}
	Among the three heatmaps: 
	while sparse sampling over short footage (left) significantly differs from dense sampling of long footage (right), 
	sparse sampling of long footage (middle) is almost equivalent to the right.	
	Video: \tucson{} (see Table~\ref{tab:videos}).}
	\vspace{-5pt}
	\label{fig:hm-time}
\end{figure}

%% file: fig-op-spd-acc.tex

\begin{figure}[t]
	\centering
	\includegraphics[width=0.35\textwidth]{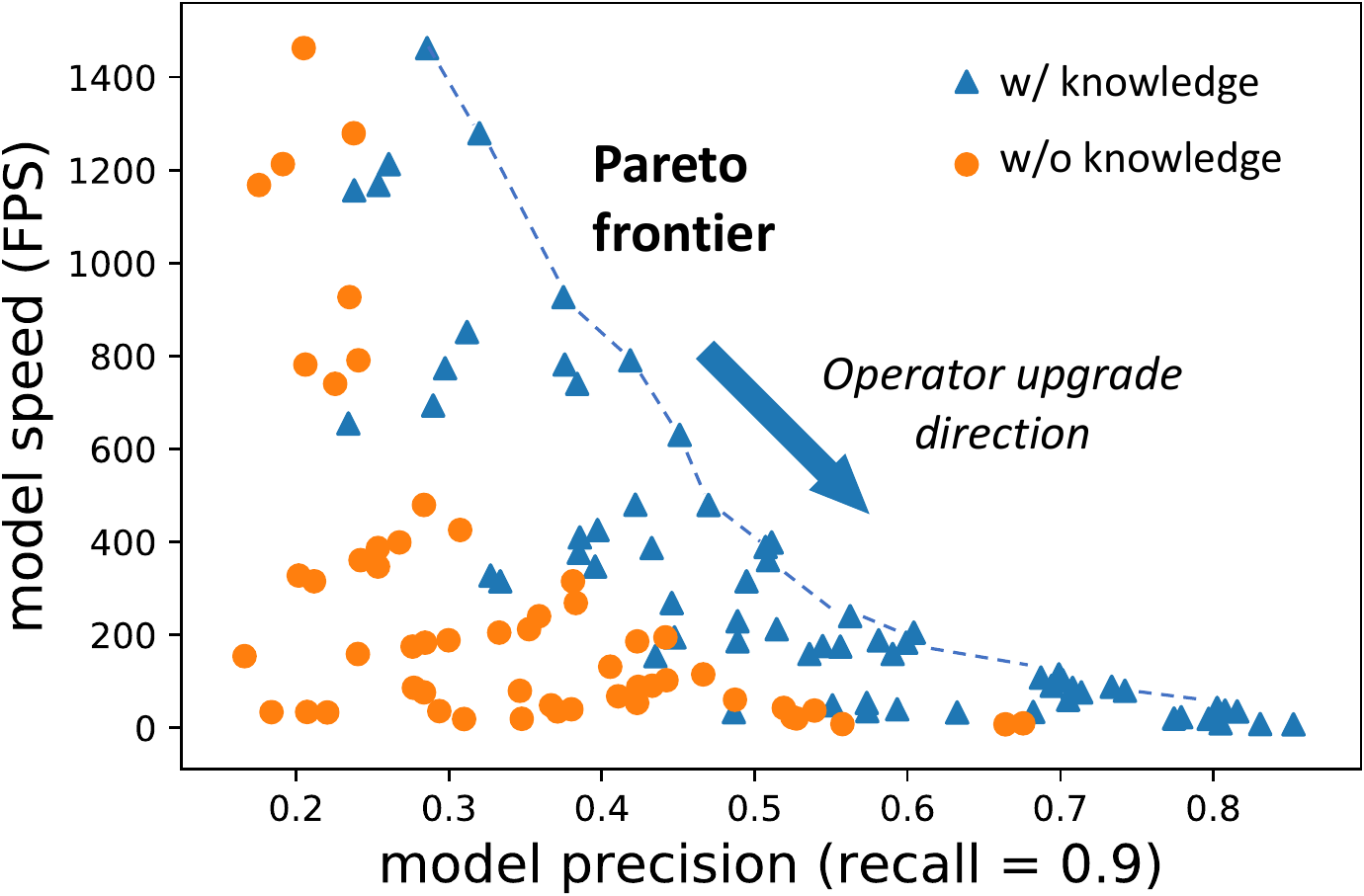}
	\caption{
	\textbf{On-camera operators benefit from long-term video knowledge substantially}. 
	Each marker: an operator.
	For querying buses on video \banff{} (see Table~\ref{tab:videos}).
	}
	\label{fig:op-spd-acc}
	\vspace{-5pt}		
\end{figure}

%% file: op.tex
\section{Online Operator Upgrade}
\label{sec:op}



\subsection{The rationale}

Three factors determine a query's execution speed: 

\begin{myenumerate}
\item 
\textit{Pending workloads}:
the difficulty of the frames to be processed, i.e., how likely will the frame be mis-filtered or mis-ranked on camera.

\item 
\textit{Camera operators}:
cheap operators spend less time on each frame but are more likely to mis-filter/mis-rank frames, especially difficult frames. This is shown in Figure~\ref{fig:op-spd-acc}.

\item 
\textit{Network condition}:
the available uplink bandwidth.
\end{myenumerate}



\noindent The three factors interplay as follows. 

\noindent $\bullet$
\textbf{Queries executed with on-camera \textit{rankers}} \hspace{1mm}
A camera ranks and uploads frames asynchronously (\S\ref{sec:overview}). 
The key is to maximize the rate of true positive frames arriving at the cloud, for which the system must balance ranking speed/accuracy with upload bandwidth.
(1) When the camera runs a \textit{cheaper} ranker, 
it ranks frames at a much higher rate than uploading the frames;
as a result, 
the cloud receives frames \textit{hastily} selected from a \textit{wide} video timespan.   
(2) When the camera runs an \textit{expensive} ranker, 
the cloud receives frames selected \textit{deliberately} from a \textit{narrow} timespan.
(3) The camera should never run rankers slower than upload, 
which is as bad as uploading unranked frames.

As an example, \OpF{} and \OpS{} on the top of Figure~\ref{fig:ranking} compare two possible executions of the same query, running cheap/expensive rankers respectively.
Shortly after the query starts (\circled{1}), \OpF{} swiftly explores more frames on camera; 
it outperforms \OpS{} by discovering and returning more true positive frames. 
As both executions proceed to harder frames (\circled{2}), 
\OpF{} makes more mistakes in ranking;
it uploads an increasingly large ratio of negatives which wastes the execution time. 
By contrast, \OpS{} ranks frames slower yet with much fewer mistakes, 
hence uploading fewer negatives. 
It eventually returns all positives earlier than \OpF{} (\circled{3}). 

The microbenchmark in Figure~\ref{fig:op-upgrade-benchmark}(a) offers quantitative evidence.
E1 spends \textit{less} time (0.7$\times$) in returning 
the first 90\% positives,
but \textit{more} time (1.7$\times$) in returning 99\% positives. 
Furthermore, \textit{lower} upload bandwidth favors 
a more \textit{expensive} ranker, as
the uploaded frames would contain a \textit{higher} ratio of positives, better utilizing the precious bandwidth.


\input{fig-ranking}
\input{fig-motivate-op-upgrade}

\noindent $\bullet$
\textbf{Queries executed with on-camera \textit{filters}} \hspace{1mm}
The key is to maximize the rate of resolving frames on camera. 
Cheap filters excel on easy frames,
resolving these frames fast with confidence.
They are incapable on difficult frames, 
wasting time on attempting frames without much success in resolving. 
They would underperform \textit{expensive} filters that spend more time per frame yet being able to resolve more frames. 


The benchmark in Figure~\ref{fig:op-upgrade-benchmark}(b) shows 
two executions with cheap/expensive filters.
Early in the query,
E1 makes faster progress as
the camera quickly resolves 50\% of the frames (4$\times$ less time than E2).
Later in the execution, E1 lags behind as
the camera cannot resolve the remaining frames and must upload them.
By contrast, E2 resolves 82\% of frames on camera and only uploads the remaining 18\%. 
As a result, 
E2 takes 1.3$\times$ less time in completing 90\% and 99\% of the query. 

\paragraph{Summary \& implications}
It is crucial for \sys{} to pick operators with optimal cost/accuracy at query time. 
The choice not only varies across queries but also varies throughout a query's execution:
easy frames are processed early, leaving increasingly difficult frames that call for more expensive operators. 
\sys{} should monitor pending frame difficulty and network bandwidth and 
upgrade operators accordingly. 





\subsection{Multipass, multi-operator execution}

\sys{} manages operators with the following techniques. 
\begin{enumerateinline}
\item 
A camera processes frames iteratively with multiple operators.
\item 
The cloud progressively updates operators on camera,
from cheaper ones to more expensive ones, 
as the direction shown in Figure~\ref{fig:op-spd-acc}. 
In picking operators, the cloud dynamically adapts operator speed to frame upload speed.
\item 
The cloud uses frames received in early execution stages to train operators for later stages;
as the latter operators are more expensive, they require more training samples. 
\end{enumerateinline}
%
%

\noindent $\bullet$ \textbfit{Multipass ranking}
This is exemplified by the bottom execution in Figure~\ref{fig:ranking}. 
The camera first runs a cheap ranker, 
moving positives towards the front of the upload queue (\circled{4}). 
Subsequently, the camera runs an expensive ranker, continuously reordering unsent frames in a smaller scope (\circled{5}). 
Throughout the query, the camera first quickly uploads easy frames that are quickly ranked and slows down to vet difficult frames with expensive/accurate ranking. 
Notably, the cheaper ranker roughly prioritizes the frames as input for the expensive ranker, ensuring the efficacy of the expensive ranker. 
In actual query executions, a camera switches among 4--8 operators
(\S\ref{sec:eval}). 

%

\noindent $\bullet$ \textbfit{Multipass filtering}
The camera sifts undecided, unsent frames in multiple passes, 
each with a more expensive filter over a sample of the remaining frames. 
Throughout one query, early, cheaper filters quickly filter easier frames, leaving more difficult frames for 
subsequent filters to resolve. 

%% file: fig-ranking.tex
\begin{figure}
	\centering
	\includegraphics[width=0.48\textwidth]{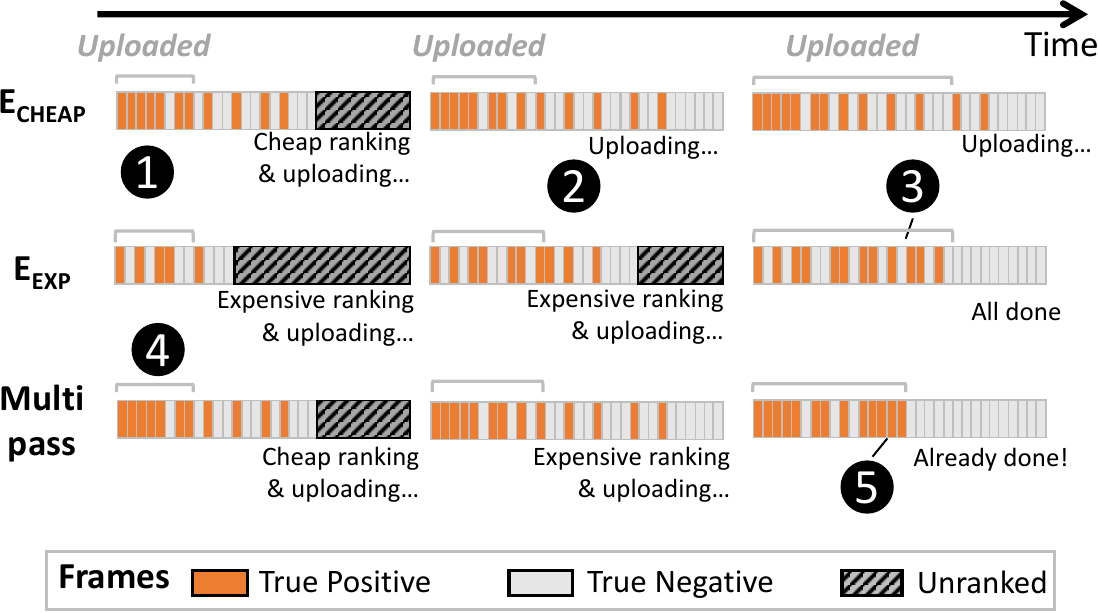}
	\caption{
	\textbf{Three alternative executions of a Retrieval query, showing multipass ranking (bottom) outperforms running individual rankers alone (top two)}. 
	Each row: 
	snapshots of the upload queue at three different moments. 
	In a queue: ranking/uploading frames from left to right. 
	}
	\label{fig:ranking}
\end{figure}

%% file: fig-motivate-op-upgrade.tex






\begin{figure}[t]
	\centering
	\includegraphics[width=0.49\textwidth]{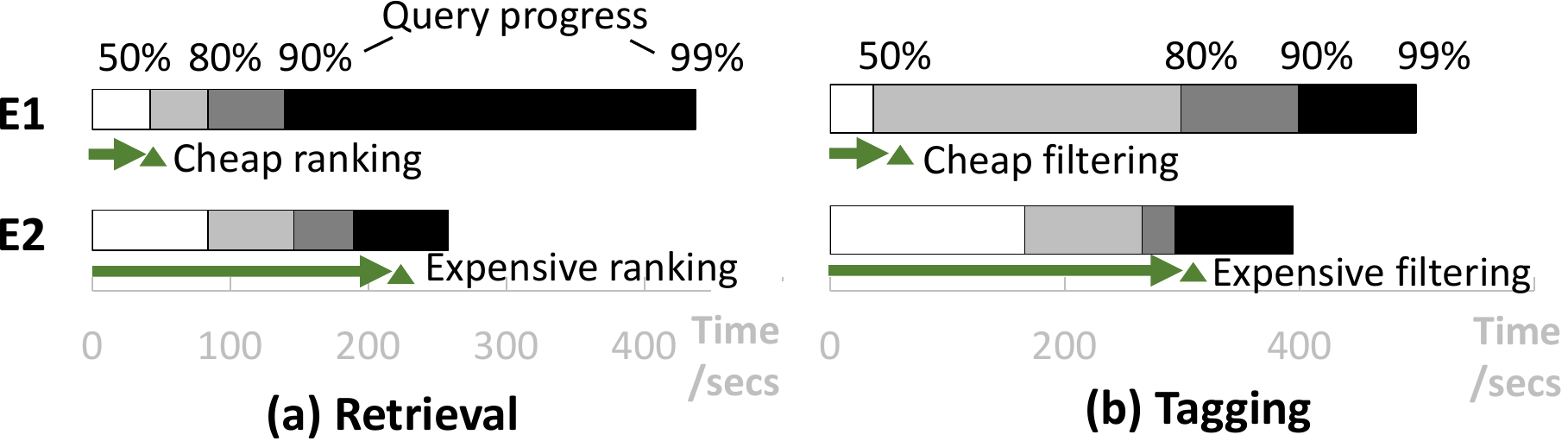}
	\caption{\textbf{Cheap/expensive camera operators excel at different query stages}. 
	Each subfigure: two alternative executions of the same query, 
	showing query progress (bars) and the corresponding operator's progress (arrows).
	}
	\label{fig:op-upgrade-benchmark}
	\vspace{-5pt}		
\end{figure}

%% file: design.tex

\section{Query Execution Planning}
\label{sec:design}
\sys{} plans a concrete query execution by 
(1) the camera's policy for selecting frames to process;
(2) the cloud's policy for upgrading on-camera operators.  
We now discuss them.
\input{design-retrieval}
\input{design-tagging}

\input{design-counting}

%% file: design-retrieval.tex
\subsection{Executing \textit{Retrieval} queries}
\label{sec:design:retrieval}

\paragraph{Policy for selecting frames}
To execute the initial operator, the camera prioritizes fixed-length video spans (e.g., 1 hour) likely rich in positive frames, estimated based on landmark frames. 
In executing subsequent operators, the camera processes frames in their existing ranking as decided by earlier operators, as described in \sect{op}.
The camera gives opportunities to frames never ranked by prior operators, 
interleaving their processing with ranked frames with mediocre scores (0.5). 




\paragraph{Policy for operator upgrade}
As discussed in \sect{overview}, 
\sys{} switches from cheap operators to expensive ones, and matches operator speed to frame upload rate. 
To capture an operator $op$'s \textit{relative} speed to upload, it uses one simple metric:
the ratio between the two speeds, i.e., $f_{op}$ = $FPS_{op} / FPS_{net}$. 
Operators with higher $f_{op}$ tend to rapidly \textit{explore} frames while others tend to \textit{exploit} slowly. 
The operator speed $FPS_{op}$ is profiled offline.
\textbfit{(1) Selecting the initial operator}
In general, \sys{} should fully utilize the upload bandwidth with positive frames. 
As positive frames are scattered in the queried video initially, 
the camera should explore all frames sufficiently fast.
Otherwise, it would either starve the uplink or 
knowingly upload negative frames. 
Based on this idea, the cloud picks the most accurate operator from the ones that are fast enough, i.e., 
$f_{op}\times R_{pos} > 1$,
%
%
where $R_{pos}$ is the ratio of positives in the queried video, estimated from landmarks.
\textbfit{(2) When to upgrade: current operator losing its vigor} 
The cloud upgrades operators either when the current operator finishes processing all frames,  
or the cloud observes a continuous quality decline in recently uploaded frames, an indicator of the current operator's incapability. 
To decide the latter, \sys{} employs a rule: the positive ratio in recently uploaded frames are $k\times$ (default: 5) lower than the frames uploaded at the beginning;
\textbfit{(3) Selecting the next operator: slow down exponentially} 
Since the initial operator promotes many positives towards the front of the upload queue, 
subsequent operators, scanning from the queue front, likely operate on a larger fraction of positives. 
Accordingly, the cloud picks the most accurate operator among much slower ones, s.t. $f_{op(i+1)} > \alpha \times f_{op(i)}$,
where $\alpha$  controls speed decay in subsequent operators. 
A larger $\alpha$ leads to more aggressive upgrade: losing more speed for higher accuracy. 
In the current prototype, we empirically choose $\alpha=0.5$. 
Since $f$ is relative to $FPS_{net}$ measured at every upgrade, 
the upgrade adapts to network bandwidth change \textit{during} a query.

%% file: design-tagging.tex
\subsection{Executing \textit{Tagging} queries}
\label{sec:design:tagging}
Recall that for \textbf{\textit{Tagging}}, a camera runs multipass filtering;
the objective of each pass is to tag, as positive ($P$) or negative ($N$), at least one frame from every K adjacent frames. 
We call $K$ the group size;
\sys{} pre-defines a sequence of group sizes as refinement levels, e.g., $K$ = 30, 10, ..., 1.
As in prior work~\cite{noscope,focus,blazeit}, the user specifies tolerable error as part of her query, e.g., 1\% false negative and 1\% false positive; \sys{} trains filters with thresholds to meet the accuracy. 

\paragraph{Policy for selecting frames}
The goal is to quickly tag easy frames in individual groups 
while balancing the workloads of on-camera processing and frame upload. 
An operator $op$ works in two stages of each pass.
i) \textit{Rapid attempting}. 
$op$ scans all the groups;
it attempts one frame per group; 
if it succeeds, it moves to the next group; it adds undecidable frames ($U$) to the upload queue.
ii) \textit{Work stealing}.
$op$ steals work from the end of upload queue. 
For an undecidable frame $f$ belonging to a group $g$, 
$op$ attempts other untagged frames in $g$; 
once it succeeds, it removes $f$ from the upload queue as $f$ no longer needs tagging in the current pass. 
After one pass, the camera switches to the next refinement level (e.g., 10 $\rightarrow$ 5). 
It keeps all the tagging results ($P$,$N$,$U$) while cancels all pending uploads. 
It re-runs the frame scheduling algorithm until it meets the finest refinement level or query terminated.

\paragraph{Policy for operator upgrade}
Given an operator $op$ and $\gamma_{op}$, the ratio of frames it can successfully tag,
\sys{} measures $op$'s efficiency by its effective tagging rate, $FPS_{op} \times \gamma_{op} + FPS_{net}$, as a sum of $op$'s successful tagging rate and the uploading rate. 
As part of operator training, 
the cloud estimates $\gamma_{op}$ for all the candidate operators by testing them 
on all landmarks (early in query) and uploaded frames (later in query). 
To select every operator, initial or subsequent, the cloud picks the candidate with the highest effective tagging rate. 
The cloud upgrades operators either when the current operator has attempted all remaining frames or another candidate having an effective tagging rate $\beta \times$ or higher (default value 2).

%% file: design-counting.tex
\subsection{Executing \textit{Counting} queries}
\label{sec:design:counting}

\paragraph{Max Count: Policy for selecting frames}
To execute the initial operator, the camera randomly selects frames to process, avoiding the worst cases that the max resides at the end of the query range. 
For subsequent operators, the camera processes frames in existing ranking decided by earlier operators.


\paragraph{Max Count: Policy for operator upgrade}
As the camera runs rankers, the policy is similar to that for \textit{Retrieval} 
with a subtle yet essential difference. 
To determine whether the current operator shall be replaced, the cloud must assess the quality of recently uploaded frames. 
While for \textit{Retrieval}, \sys{} conveniently measures the quality as the ratio of \textit{positive} frames, 
the metric does not apply to \textit{\countmax}, which seeks to discover \textit{higher} scored frames. 
Hence, \sys{} adopts the Manhattan distance as a quality metric among the permutations from the ranking of the uploaded frames (as produced by the on-camera operator) and the ranking that is re-computed by the cloud object detector. 
A higher metric indicates worse quality hence more urgency for the upgrade.

\paragraph{Average/Median Count: no on-camera operators}
After the initial upload of landmarks, the camera randomly samples frames in queried videos and uploads them for the cloud to refine the average/median statistics. 
To avoid any sampling bias, the camera does not prioritize frames; 
it instead relies on the Law of Large Numbers (LLN)~\cite{encyclopaediaofmathematics} to approach the average/median ground truth through continuous sampling. 

%% file: impl.tex





 
\section{Implementation and Methodology}
\label{sec:impl}

\paragraph{Operators}
We architect on-camera operators as variants of AlexNet~\cite{alexnet}. 
We vary the number of convolutional layers (2--5), convolution kernel sizes (8/16/32), 
the last dense layer's size (16/32/64); 
and the input image size (25$\times$25/50$\times$50/100$\times$100).
We empirically select 40 operators to be trained by \sys{} online; 
we have attempted more but see diminishing returns.
These operators require small training samples (e.g., 10K images) and run fast on camera.
\noindent
\textbf{Background subtraction} filters static frames at low overhead~\cite{focus}.
\sys{} employs a standard technique~\cite{bng-sub}:
during video capture, a camera detects frames that have little motion (< 1\% foreground mask) and omits them in query execution.
On our camera hardware (Table~\ref{tab:exp}), background subtraction is affordable in real time during capture. 
For fair comparisons, we augment all baselines with background subtraction. 

\input{eval-method}

%% file: eval-method.tex
\input{tab-hwmodel}
\input{tab-videos}

\paragraph{Videos \& Queries}
\revision{
We test \sys{} on 15 videos captured from 15 live camera feeds (Table~\ref{tab:videos}). 
Each video lasts continuous 48 hours including daytime and nighttime, collected between Oct. 2018 to Mar. 2019.
We preprocess all videos to be 720P at 1 FPS, consistent with prior work~\cite{focus}.
We test \textit{Retrieval}/\textit{Tagging}/\textit{Counting} queries on 6/6/3 videos. 
We intentionally choose videos with disparate characteristics and hence different degrees of difficulty.
For instance, 
\whitebay{} is captured from a close-up camera, containing clear and large persons;
\venice{} is captured from a high camera view and hence contains blurry and small persons. 
For each video, we pick a representative object class to query; 
across videos, these classes are diverse.
}
For \textit{Tagging}, we set query error to be $<$ 1\% FN/FP as prior work did~\cite{noscope}.

\revision{A query's accuracy is reflected by its execution progress.
For retrieval/counting, we report accuracy as the fraction of positive frames returned. There is no false positive because the cloud always runs the high-accuracy object detector as a ``safety net'', of which the output is regarded as the ground truth.
For tagging, we report accuracy as query errors, meaning the percentage of frames mistakenly tagged. To issue a query, the user sets the target error, which by default is 1\% as in prior work. Table~\ref{tab:query} and $\S$\ref{sec:design} provide more details.
}



\paragraph{Test platform \& parameters}
As summarized in Table~\ref{tab:exp}(a), 
we test on embedded hardware similar to low-cost cameras~\cite{cheapcam,wyzecam}. 
We use Rpi3 as the default camera hardware and report its measurement 
unless stated otherwise.
During query execution, both devices set up a network connection with 1MB/s default bandwidth to emulate typical WiFi condition~\cite{mpdash}.
Note that this network bandwidth is \textit{not} meant for streaming; it is only for a camera while the camera is being queried. 
We run YOLOv3 as the high-accuracy object detector on camera and cloud (Table~\ref{tab:exp}(b)). 
\revision{In calculating accuracy, we use the output of YOLOv3 as the ground truth as in prior work~\cite{noscope,focus}.
On Rpi3, we partition YOLOv3 into three stages, each fitting into DRAM separately.
}
We will study alternative models, landmarks, and resources in \sect{eval:lm}.


\paragraph{Baselines}
As summarized in Table~\ref{tab:exp}(b), we compare \sys{} with three alternative designs augmented with background subtraction and only process/transmit non-static video frames. 
%
%
\begin{myitemize}
\item 
\textbf{\cloudonly{}}: a naive design that uploads all queried frames at query time for the cloud to process.

\item
\textbf{\noscope{}}: in the spirit of NoScope~\cite{noscope}, 
the camera runs only one ranker/filter specialized for a given query, selected by a cost model for minimizing full-query delay.
To make \noscope{} competitive, we augment it with landmark frames to reduce the operator training cost.
Compared to \sys{}, \noscope{}'s key differences are the lack of operator upgrade and 
the lack of operator optimization by long-term video knowledge.
%
\item 
\textbf{\focus{}}:
in the spirit of Focus~\cite{focus}, the camera runs a cheap yet generic object detector on all frames. 
We pick YOLOv3-tiny (much cheaper than YOLOv3) as the detector affordable by Rpi3 in real time (1 FPS).
The detector plays the same role as an operator in \sys{}, except that it runs at capture time:
for \textit{Retrieval} and \textit{Counting}, the detector's output scores are used to prioritize frames to upload at query time;
for \textit{Tagging}, the output is used to filter the frames that have enough confidence.
\revision{\focus{} implements all run-time features of Focus except feature clustering. We left out clustering because we find it performs poorly on counting queries.}
Compared to \sys{}, \focus{}'s key differences are: it answers queries solely based on the indexes built at capture time; 
it requires no operator training or processing actual images at query time. 


\end{myitemize}

%% file: tab-hwmodel.tex
\begin{table}[t!]

\begin{subtable}[t]{0.47\textwidth}
\centering
\includegraphics[width=1\textwidth{}]{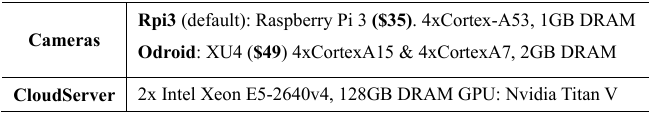} 
\vspace{-25pt}
\caption{\textit{Hardware platforms}}
\end{subtable}


\begin{subtable}[t]{0.5\textwidth}
\scriptsize
\begin{tabular}{l|l|l|c}
\hline
\textbf{} & \textbf{Cam:Landmarks} & \textbf{Cam:Query} & \textbf{Cloud:Query} \\ 
\hline
ClondOnly 	& -- & Only upload frames &  \\ 
OptOp  	& Yv3 every 30 secs & Run one optimal op & Yv3 on all \\ 
PreIndexAll & YTiny every sec & Parse YTiny result & uploaded frames \\ 
\sys{}    & Yv3 every 30 secs & Multi passes \& ops &  \\ \hline
\end{tabular}
\caption{
\textit{\sys{} and the baselines}. The table summarizes their executions for capture and query. 
NNs: Yv3 -- YOLOv3, high accuracy (mAP=57.9); 
YTiny -- YOLOv3-tiny, low accuracy (mAP=33.1). 
}
\end{subtable}
\vspace{-10pt}
\caption{\textbf{Experiment configurations}}
\label{tab:exp}
\vspace{-5pt}
\end{table}


%% file: tab-videos.tex

\begin{table}[]
\scriptsize
\begin{tabular}{c | l l l}
\hline
\textbf{} & \textbf{Name} & \textbf{Object} & \textbf{Description} \\ \hline
\multirow{8}{*}{T} & \jacksonh~\cite{JacksonH} & car & A busy intersection in Jackson Hole, WY \\ \cline{2-4} 
 & \jacksont{}~\cite{JacksonT} & car & A night street in Jackson Hole, WY \\ \cline{2-4} 
 & \banff{}~\cite{Banff} & bus & A cross-road in Banff, Alberta, Canada \\ \cline{2-4} 
 & \mierlo{}~\cite{Mierlo} & truck & A rail crossing in Netherlands \\ \cline{2-4} 
 & \miami~\cite{Miami} & car & A cross-road in Miami Beach, FL \\ \cline{2-4} 
 & \ashland~\cite{Ashland} & train & A level crossing in Ashland, VA \\ \cline{2-4}
 & \shibuya~\cite{Shibuya} & bus & An intersection in Shibuya (\begin{CJK}{UTF8}{min}渋谷\end{CJK} ), Japan \\ \hline
\multirow{5}{*}{\begin{tabular}[c]{@{}l@{}}O\end{tabular}} & \chaweng~\cite{Chaweng} & bicycle & Absolut Ice Bar (outside) in Thailand \\ \cline{2-4} 
 & \lausanne~\cite{Lausanne} & car & A pedestrian plaza in Lausanne, Switzerland \\ \cline{2-4} 
 & \venice~\cite{Venice} & person & A waterfront walkway in Venice, Italy \\ \cline{2-4} 
 & \oxford~\cite{Oxford} & bus & A street beside Oxford Martin school, UK \\ \cline{2-4}
 & \whitebay~\cite{Whitebay} & person & A beach in Virgin Islands \\ \hline
\multirow{2}{*}{\begin{tabular}[c]{@{}l@{}}I\end{tabular}} & \coralreef~\cite{CoralReef} & person & An aquarium video from CA \\ \cline{2-4} 
 & \boathouse~\cite{BoatHouse} & person & A retail store from Jackson Hole, WY \\ \hline
\multirow{1}{*}{W} & \eagle~\cite{Eagle} & eagle & A tree with an eagle nest in FL \\ \hline
\end{tabular}
\vspace{-5pt}
\caption{\textbf{15 videos used for test}. 
	Each video: 720P at 1FPS lasting 48 hours.
	Column 1: video type. T -- traffic; O/I -- outdoor/indoor surveillance; W -- wildlife.}
\vspace{-5pt}
\label{tab:videos}
\end{table}


%% file: eval.tex

\section{Evaluation}
\label{sec:eval}




\input{eval-perf}
\input{eval-comp}

\input{eval-local}

%% file: eval-perf.tex
\subsection{End-to-end performance}
\label{sec:eval:e2e}


\input{fig-perf-combine}
\input{fig-counting-max}

\paragraph{Full query delay}
is measured as:
(\textit{Retrieval}) the time to receive 99\% positive frames
as in prior work~\cite{focus};
(\textit{Tagging}) the time taken to tag every frame;
(\textit{Counting}) the time to reach the ground truth (max) or converge within 1\% error of the ground truth (avg/median). 
Overall, \sys{} delivers good performance and outperforms the baselines significantly.

\begin{myitemize}
\item 
\textit{Retrieval} (Figure~\ref{fig:perf-combine}(a)). On videos each lasting 48 hours, \sys{} spends $\sim$1,900 seconds on average, i.e., 89$\times$ of video realtime.
On average, \sys{}'s delay is 3.8$\times$, 3.1$\times$, and 2.0$\times$ shorter than \cloudonly{}, \focus{}, and \noscope{}, respectively.

\item 
\textit{Tagging} (Figure~\ref{fig:perf-combine}(b)). 
\sys{} spends $\sim$581 seconds on average (297$\times$ realtime).
This delay is 16.0$\times$, 2.1$\times$, and 4.3$\times$ shorter than \cloudonly{}, \focus{}, and \noscope{}, respectively. 

\item 
\textit{Counting} (Figure~\ref{fig:counting}).
\sys{}'s average/median take several seconds to converge.
For \textit{average Count}, 
\sys{}'s delay is 65.1$\times$ and up to three orders of magnitude shorter than \cloudonly{} and \focus{}. 
For \textit{median Count}, \sys{}'s delay is 68.3$\times$  shorter than the others. 
For \textit{\countmax}, \sys{} spends 34 seconds on average (635$\times$ realtime), which is 5.8$\times$, 5.0$\times$, and 1.3$\times$ shorter than \cloudonly{}, \focus{}, and \noscope{}.
\end{myitemize}

\paragraph{Query progress}
\sys{} makes much faster progress in most time of query execution.
It \textit{always} outperforms \cloudonly{} and \noscope{}
during \textit{Retrieval}/\textit{Tagging} (Figure~\ref{fig:perf-combine}).
It \textit{always} outperforms alternatives in median/average count (Figure~\ref{fig:counting}).

\paragraph{Why \sys{} outperforms the alternatives?}
The alternatives suffer from the following.
\begin{myenumerateinline}
\item 
Inaccurate indexes. 
\focus{} resorts to inaccurate indexes (YOLOv3-tiny) built at capture time.
Misled by them, \textit{Retrieval} and \textit{Tagging} upload too much garbage;
\textit{Counting} includes significant errors in the initial estimation, slowing down convergence. 

\item 
Lack of long-term knowledge. 
\noscope{}'s operators are either slower or less accurate than \sys{}, as illustrated in Figure~\ref{fig:op-spd-acc}.

\item 
One operator does not fit an entire query. 
Optimal at some point (e.g., 99\% Retrieval), the operator runs too slow on easy frames which could have been done by cheaper operators.

\end{myenumerateinline}




\paragraph{Why \sys{} underperforms (occasionally)?}
On short occasions, 
\sys{} may underperform \focus{} at early query stages, e.g., \boathouse{} in Figure~\ref{fig:perf-combine}. 
Reasons:
(1) \focus{}'s inaccurate indexes may be correct on \textit{easy} frames;
(2) \focus{} does not pay for operator bootstrapping as \sys{}. 
Nevertheless, \focus{}'s advantage is transient.
As easy frames are exhausted, indexes make more mistakes on the remaining frames and hence slow down the query.

\input{tab-bw}

\paragraph{Can \sys{} outperform under different network bandwidths at query time?}
Table~\ref{tab:bw} summarizes \sys{}'s query delays at 9 bandwidths evenly spaced in [0.1 MB/s, 10 MB/s] which cover typical WiFi bandwidths~\cite{wifi-state}.
We have observed that: on lower bandwidths, \sys{}'s advantages over baselines are more significant; at high bandwidths, \sys{}'s advantages are still substantial (>2$\times$ in most cases) yet less pronounced. 
The limitation is not in \sys{}'s design but rather its current NNs: 
we find it difficult to train operators that are both fast enough to keep up with higher upload bandwidth and accurate enough to increase the uploaded positive ratio proportionally.

\paragraph{vs. ``all streaming'': query speed}
As ``all streaming'' uploads all videos to the cloud before a query starts, 
the query speed is bound by cloud GPUs but not network bandwidth.
With our default experiment setting (1 GPU and 1MB/s network bandwidth), ``all streaming'' still runs queries much slower than zero streaming.
Adding more cloud GPUs will eventually make ``all streaming'' run faster than \sys{}. 

\input{fig-net-save}

\paragraph{vs. ``all streaming'': network bandwidth saving}
Compared to  streaming all videos (720P 1FPS) at capture time,
\sys{} saves traffic significantly, as shown in Figure~\ref{fig:eval-net-save}. 
When only as few as 0.005\% of video is queried as in our case study (\S\ref{sec:bkgnd}), the saving is over three orders of magnitude.
Even if all captured videos are queried, \sys{} saves more than 10$\times$, 
as its on-camera operators skip uploading many frames. 
Among the bandwidth reduction brought by \sys, only less than 30\% attributes to the background subtraction technique.
It shows that the disadvantage of ``all streaming'' is fundamental: streaming optimizations may help save the bandwidth (upmost several times~\cite{vstore}) but cannot offset the waste, as discussed in \sect{bkgnd:case}.

\input{eval-train}

%% file: fig-perf-combine.tex
\begin{figure*}[ht]
	\centering					
	\begin{minipage}[b]{0.96\textwidth}
		\includegraphics[width=1\textwidth]{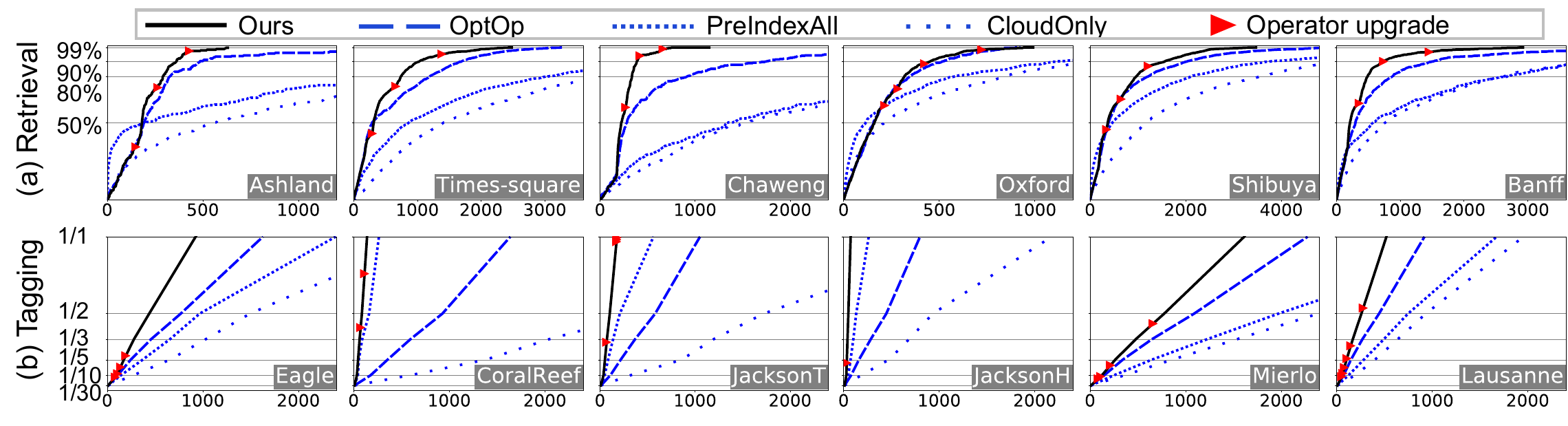}
	\end{minipage}
	\vspace{-10pt}
	\caption{\textbf{On \textit{Retrieval} and \textit{Tagging} queries, \sys{} shows good performance and outperforms the alternatives}.
	x-axis for all: query delay (secs).
	y-axis for (a): \% of retrieved instances;
	for (b): refinement level (1/N frames).
	}
	\vspace{-10pt}
	\label{fig:perf-combine}
\end{figure*}

	

	

%

%% file: fig-counting-max.tex

\begin{figure}[t]
	\centering					
	\includegraphics[width=.42\textwidth]{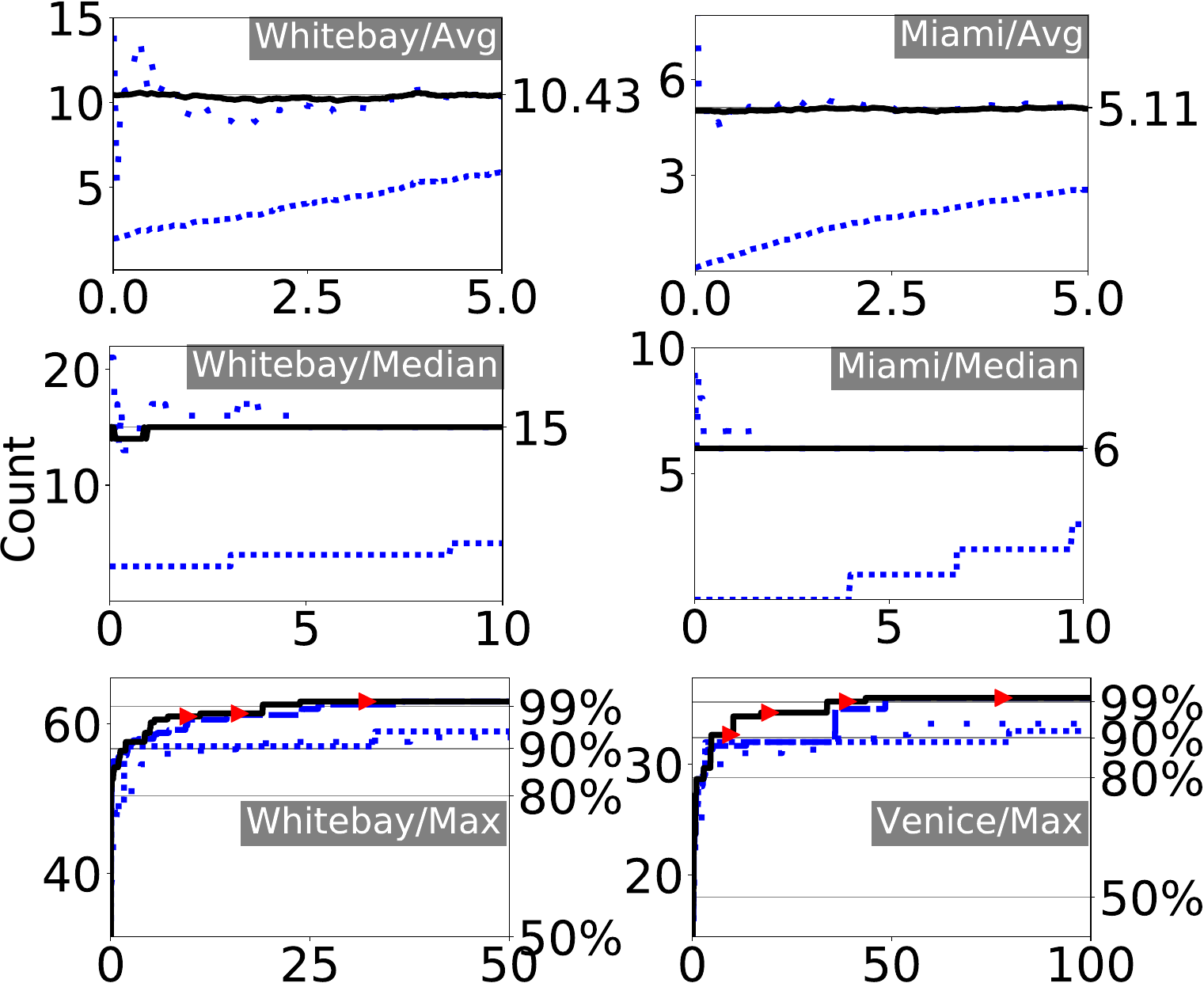}
	\caption{ \textbf{On \textit{Counting} queries, \sys{} shows good performance and outperforms the alternatives}.
				Legend: see Figure~\ref{fig:perf-combine}.
				x-axis for all: query delay (secs).
				y-axis for left plots: count; 
				for top two right plots: ground truth for avg/median queries; 
				for bottom right plot: \% of max value.
				}
	\label{fig:counting}
	\vspace{-10pt}
\end{figure}

%
%
%
%
%

%% file: tab-bw.tex
\begin{table}[t]
\centering
\includegraphics[width=0.45\textwidth]{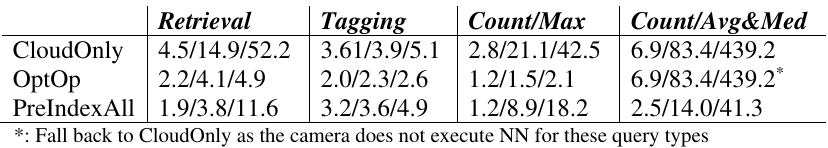}
\vspace{-5pt}
\caption{
\sys{}' performance (speedup) with various bandwidths.
Numbers: min/median/max of times ($\times$) of query delay reduction  compared to baselines (rows). 
Averaged on all videos and 9 bandwidths in 0.1MB/s--10MB/s.}
\label{tab:bw}
\vspace{-5pt}
\end{table}

%% file: fig-net-save.tex
\begin{wrapfigure}{r}{0.25\textwidth}
	\centering
	\includegraphics[width=.20\textwidth]{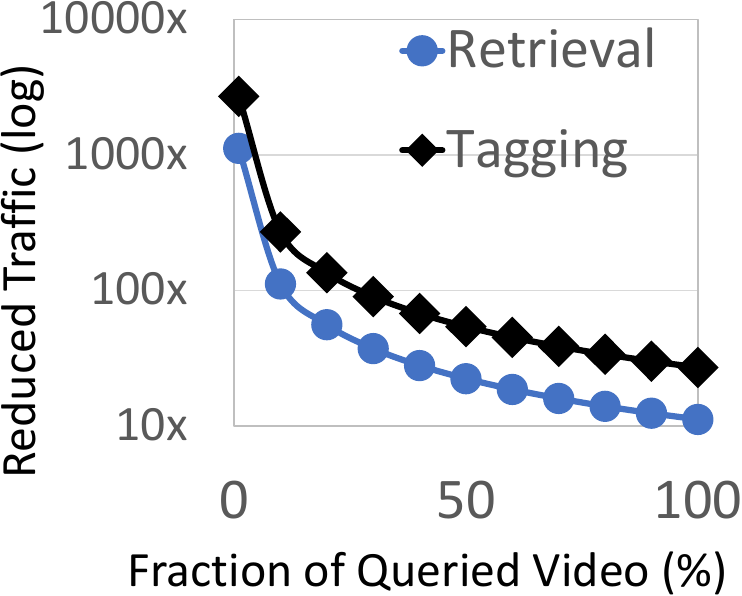}
	\vspace{-5pt}
	\caption{\textbf{\sys{} significantly reduces network traffic compared to ``all streaming''.}
	Results averaged over all videos.
	}
	\vspace{-5pt}
	\label{fig:eval-net-save}
\end{wrapfigure}

%% file: eval-train.tex
\paragraph{Training \& shipping operators}
For each query, \sys{} trains $\sim$40 operators, of which $\sim$10 are on the Pareto frontier.
The camera switches among 4--8 operators, which run at diverse speeds (27$\times$--1,000$\times$ realtime) and accuracies. 
\sys{} chooses very different operators for different queries.
Training one operator typically takes 5--45 seconds on our test platform
and requires 5k frames (for bootstrapping) to 15k frames (for stable accuracy). 
Operators' sizes range from 0.2--15 MB.
Sending an operator takes less than ten seconds.
Only the delay in training and sending the first operator ($\le$ 40 seconds) adds to the query delay which is included in Figure~\ref{fig:perf-combine}/\ref{fig:counting}. 
Subsequent operators are trained and transmitted in parallel to query execution. 
Their delays are hidden from users.

\revision{
\noindent \textbf{\sys{} elasticity}
Due to \sys{}'s design, the computing resources available on low-cost cameras are used efficiently at both capture and query time.
Thanks to its elastic execution,
it can avoid interference with a camera’s surveillance task, notably video encoding and storage. 
For instance, DIVA can produce denser/sparser landmarks per its CPU time allocated by the camera OS.
According to our experiments on Raspberry Pi 3B+, recording video at 720P and 30 FPS only uses less than 2\% of CPU time, which is negligible as compared to NN execution.
We reserve a small fraction of CPU time to surveillance using \textit{cgroup} and observe no frame drop in the surveillance task and negligible slowdown in NN execution.
}

%% file: eval-comp.tex
\subsection{Validation of query execution design}
\label{sec:eval:exec}

\input{fig-component}

The experiments above show \sys{}'s substantial advantage over \noscope{}, 
coming from a combination of two techniques -- 
optimizing queries with long-term video knowledge (``\textit{Long-term opt}'', \S\ref{sec:lm})
and operator upgrade (``\textit{Upgrade}'', \S\ref{sec:op}). 
We next break down the advantage by incrementally disabling the two techniques in \sys{}.
Figure~\ref{fig:eval-comp} shows the results. 

\noindent
\textbf{Both techniques contribute to significant performance.} 
For instance, disabling \textit{Upgrade} increases the delay of retrieving 90\% instances by 2$\times$ and that of tagging 1/1 frames by 2$\times$-3$\times$. 
Further disabling \textit{Long-term Opt} increases the delay of Retrieval by 1.3$\times$-2.1$\times$ and that of tagging by 1.6$\times$-3.1$\times$.
Both techniques disabled, \sys{} still outperforms \cloudonly{} with its single non-optimized operator.

\noindent
\textbf{\textit{Upgrade}'s benefit is universal;
\textit{Long-term opt}'s benefit is more dependent on queries}, i.e., the skews of the queried object class in videos. 	
For instance, \sys{}'s benefit is more pronounced on \chaweng{}, where small bicycles only appear in a region in 1/8 size of the entire frame, than \ashland{}, where large trains take 4/5 of the frame.
With stronger skews in \chaweng, \sys{} trains operators that are more accurate and run faster. 
This also accounts for \sys{}'s varying (yet substantial) advantages over the alternatives (Figure~\ref{fig:perf-combine}).

%% file: fig-component.tex
\begin{figure}[t]
\centering					
	\includegraphics[width=.45\textwidth]{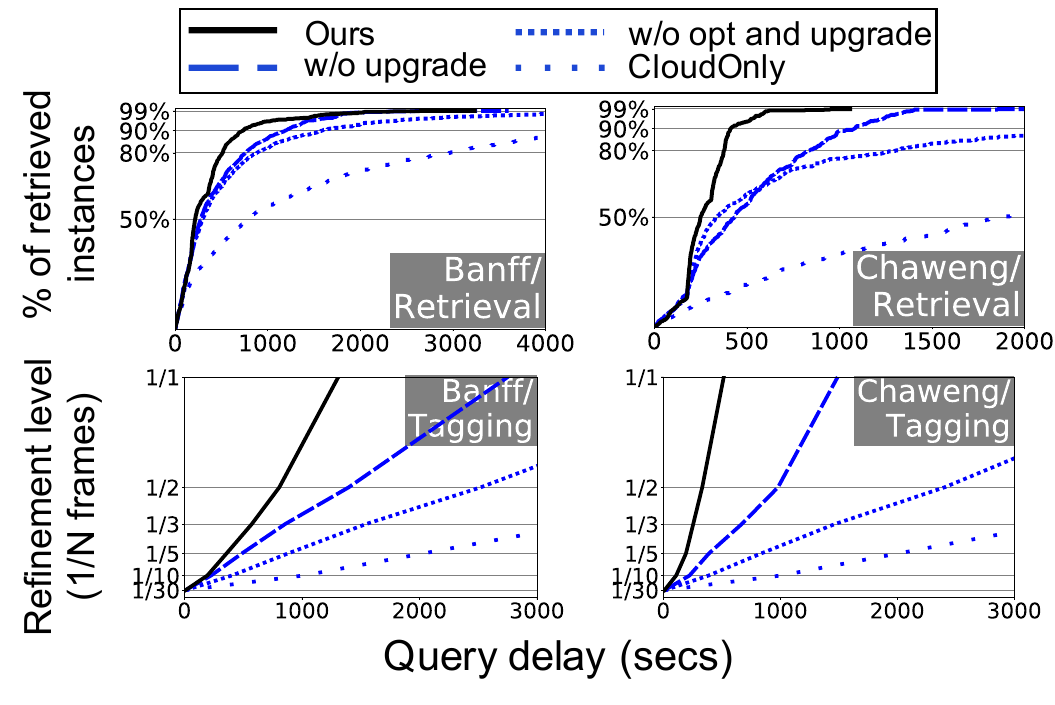}
	\vspace{-10pt}
	\caption{
	\sys{}'s both key techniques -- optimization with long-term video knowledge (opt) and operator upgrade (upgrade), contribute to performance significantly.}
	\label{fig:eval-comp}
	\vspace{-5pt}
\end{figure}




%% file: eval-local.tex
\subsection{Validation of landmark design}
\label{sec:eval:lm}



\input{fig-local-oracle}

Next, we deviate from the default landmark parameters (Table~\ref{tab:exp}) to validate the choice of sparse-but-sure landmarks.


\paragraph{\sys{} hinges on accurate landmarks.}
As shown in Figure~\ref{fig:local-oracle}(a), modestly inaccurate landmarks (as produced by YOLOv2; 48.1 mAP) increase delays for Q1/Q2 by 45\% and 17\%. 
Even less accurate landmarks (by YOLOv3-tiny; 33.1 mAP) increase the delays significantly by 5.3$\times$ and 4.3$\times$. 
Perhaps surprisingly, such inaccurate landmarks can be worse than no landmarks at all (``w/o LM'' in Figure~\ref{fig:local-oracle}): when a query starts, a camera randomly uploads unlabeled frames for the cloud to bootstrap operators.
\hspace{1mm}
\textit{Why inaccurate landmarks hurt so much?} 
They (1) provide wrong training samples; 
(2) lead to incorrect observation of spatial skews which further mislead frame cropping; 
and (3) introduce large errors into initial statistics, making convergence harder. 

\paragraph{\sys{} tolerates longer landmark intervals.}
As shown in Figure~\ref{fig:local-oracle}(b), 
\sys{}'s \textit{Retrieval} and \textit{Tagging} performance slowly degrade with longer intervals. 
Even with an infinite interval, i.e., ``w/o LM'' in Figure~\ref{fig:local-oracle}(a),
the slowdown is no more than 3$\times$.
On \textit{Counting}, 
the performance degradation is more pronounced:
5$\times$ longer intervals for around 15$\times$ slow down. 
Yet, such degradation is still much smaller than one from inaccurate landmarks (two orders of magnitude). 
The reason is that, with longer LM intervals \sys{} has to upload additional frames in full resolution ($\sim$10$\times$ larger than LMs) when a query starts for bootstrapping operators; such a one-time cost, however, is amortized over the full query.




\paragraph{Create the most accurate landmarks possible}
Should a camera build denser yet less accurate landmarks or sparser yet more accurate ones? 
Figure~\ref{fig:local-oracle}(c) suggests the latter is always preferred,
because of \sys{}'s high sensitivity to landmark accuracy and low sensitivity to long landmark intervals. 

\paragraph{\sys{} on wimpy/brawny cameras}
\sys{} suits wimpy cameras that can only generate sparse landmarks.
Some cameras may have DRAM smaller than a high-accuracy NN (e.g., $\sim$1 GB for YOLOv3); 
fortunately, recent orthogonal efforts reduce NN sizes~\cite{split-NN}.
Wimpier cameras will further disadvantage the alternatives,
e.g., \focus{} will produce even less accurate indexes. 
On higher-end cameras (a few hundred dollars each~\cite{coral-camera})
that \sys{} is \textit{not} designed for, 
\sys{} still shows benefits, albeit not as pronounced. 
High-end cameras can afford more computation at capture time. 
i) They may run \focus{} with improved index accuracy. 
In Figure~\ref{fig:local-oracle}(a), 
running YOLOv2 on all captured frames (\focus{}+Yv2), \sys{}'s performance gain is 1.9$\times$ (left) or even 0.6$\times$ (right).
ii) These cameras may generate denser landmarks and rely on the cloud for the remaining frames. 
Figure~\ref{fig:local-oracle}(b) shows, with one landmark every 5 seconds, \sys{}'s advantage is 1.5$\times$. 

%% file: fig-local-oracle.tex
\begin{figure}[t]
	\centering		
	\begin{minipage}[b]{0.4\textwidth}
		\includegraphics[width=1\textwidth]{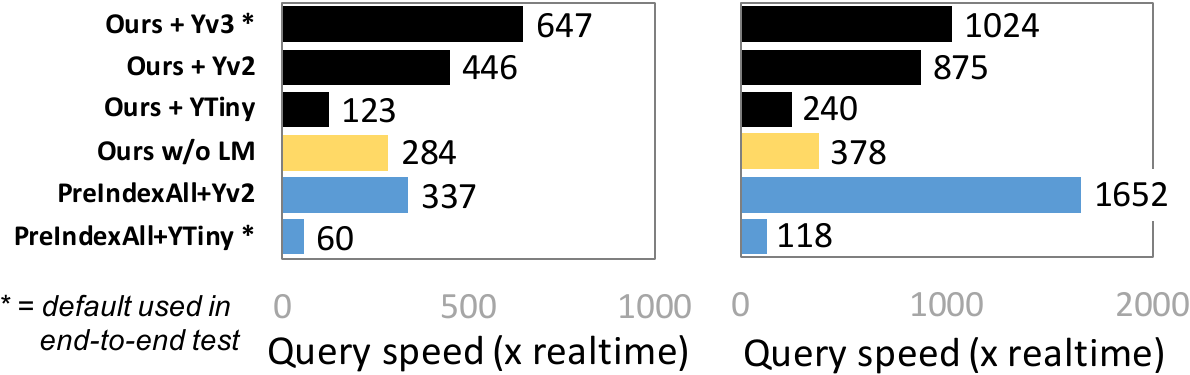}
		\subcaption{		
		\sys{}'s performance degrades significantly with less accurate landmarks (produced by Yv2 and YTiny), 
		which can be even worse than no landmarks at all (``w/o LM'').
		}
	\end{minipage}

	\centering		
	\begin{minipage}[b]{0.4\textwidth}
		\includegraphics[width=1\textwidth]{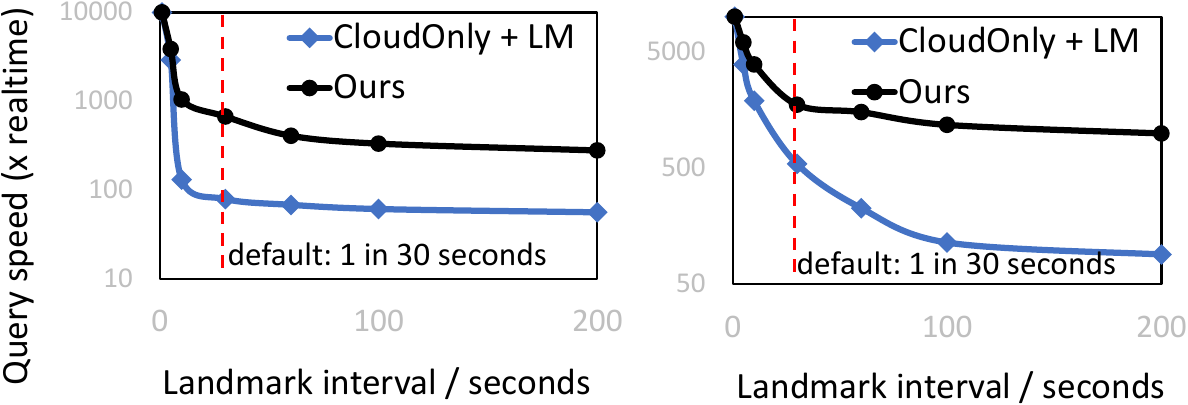}
		\subcaption{
		\sys{}'s performance degrades slowly with sparser landmarks.		
		The y-axis is logarithmic.		
		}
	\end{minipage}	

	\centering		
	\begin{minipage}[b]{0.4\textwidth}
		\includegraphics[width=1\textwidth]{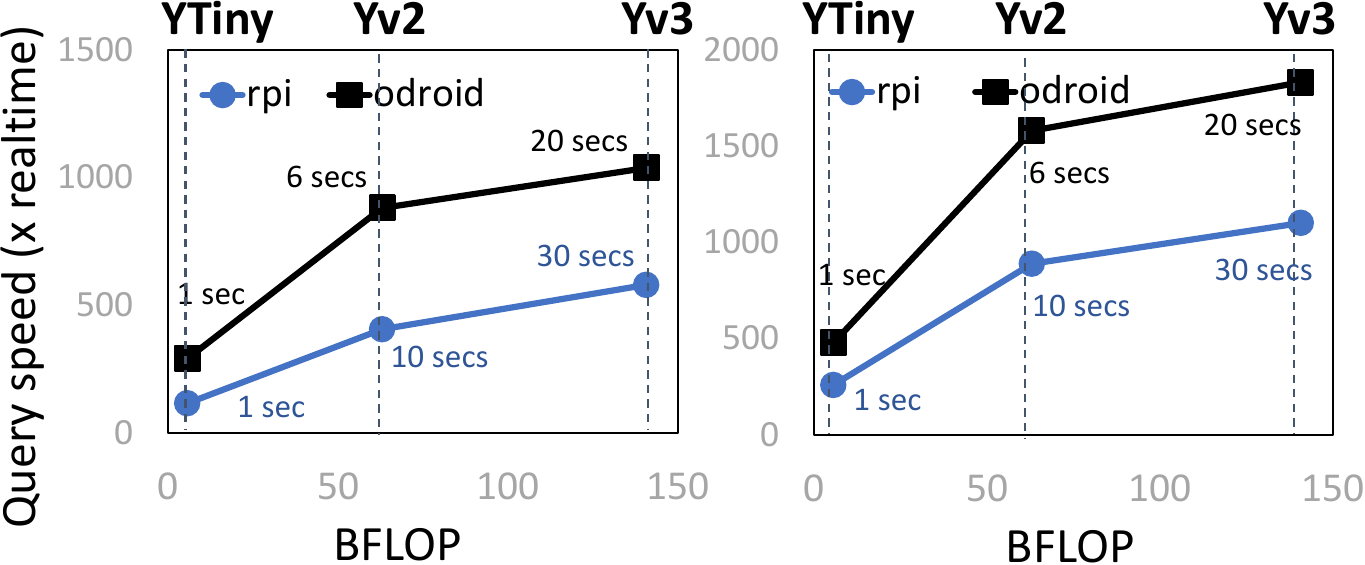}
		\subcaption{On given camera hardware (Rpi3/Odroid), 
		sparser yet more accurate LMs always improve \sys{}'s performance.
		Landmark intervals annotated along curves.
	}
	\end{minipage}

	\caption{\textbf{Validation of landmark design}. 
	In (a)/(b)/(c): Left -- \textit{Retrieval} on \chaweng; 
	Right -- \textit{Tagging} on \jacksonh. 
	}
		
	\label{fig:local-oracle}
	\vspace{-15pt}
\end{figure}

%% file: related.tex
\section{Related Work}
\label{sec:related}

%
%



\paragraph{Optimizing video analytics}
The CV community has studied video analytics for decades, 
e.g., for online training~\cite{clickbait,clickbaitv2} and active learning~\cite{activelearning}.
They mostly focus on improving analytics accuracy on short videos~\cite{saligrama12cvpr, liu17cvpr, zhu18cvpr, kang16cvpr, secs, shen17cvpr} while missing opportunities in exploiting long-term knowledge (\S\ref{sec:lm}). 
These techniques alone cannot address the systems challenges we face, e.g., network limit or frame scheduling. 
%
A common theme of recent work is to trade accuracy for lower cost:
VStore~\cite{vstore} does so for video storage;
Pakha \textit{et al.}~\cite{pakha18hotcloud} do so for network transport; 
Chameleon~\cite{chameleon} and VideoStorm~\cite{videostorm, videoedge} do so with video formats. 
\sys{}'s operators as well exploit accuracy/cost tradeoffs.
Multiple systems analyze archival videos on servers~\cite{scanner,deeplens,vstore,blazeit,shen2019nexus}. 
\sys{} analyzes archival videos on \textit{remote} cameras and embraces new techniques.
ML model cascade is commonly used for processing a stream of frames~\cite{noscope,objectdetect-cascade,pedestrian-cascade}:
in processing a frame, it keeps invoking a more expensive operator if the current operator has insufficient confidence.
This technique, however, mismatches exploratory analytics, for which \sys{} uses one operator to process many frames in one pass and produces inexact yet useful results for all of them. 

\paragraph{Edge video analytics}
To reduce cloud/edge traffic, computation is partitioned, e.g., 
between cloud/edge~\cite{lavea,deepdecision,filterforward}, edge/drone~\cite{wang18sec}, and edge/camera~\cite{vigil}. 
Elf~\cite{elfcam} executes counting queries completely on cameras.
Most work targets live analytics, processes frames in a streaming fashion and trains NNs ahead of time. 
\sys{} spreads computation between cloud/cameras but
takes a disparate design point (zero streaming) that are inadequate in prior systems.
CloudSeg~\cite{cloudseg}
reduces network traffic by uploading low-resolution frames and recovering them via super resolution. \sys{} eliminates network traffic at capture time at all. 

\paragraph{Online Query Processing}
Dated back in the 90s, online query processing allows users to see early results and control query execution~\cite{ola, control}.
It is proven effective in large data analytics, such as MapReduce~\cite{olamr}. 
\sys{} retrofits the idea for video queries and accordingly contributes new techniques, e.g., operator upgrade, to support the online fashion. 
\sys{} could borrow UI designs from existing online query engines. 




\paragraph{WAN Analytics}
To query geo-distributed data, recent proposals range from query placement to data placement~\cite{wanalytics, vulimiri15nsdi, clarinet, iridium, lube}.
JetStream~\cite{jetstream} adjusts data quality to meet network bandwidth; AWStream~\cite{awstream} facilitates apps to systematically trade-off analytics accuracy for network bandwidth. 
Like them, \sys{} adapts to network; 
unlike them, \sys{} does so by changing operator upgrade plan, a unique aspect in video analytics.
\sys{} targets resource-constrained cameras, which are unaddressed in WAN analytics.

%% file: conclusions.tex

\section{Conclusions}
\label{sec:conclusion}
Zero streaming shifts most compute from capture time to query time.
We build \sys, an analytics engine for querying cold videos on remote, low-cost cameras. 
At capture time, \sys{} builds sparse but sure landmarks; at query time, it refines query results by continuously updating on-camera operators.
Our evaluation of three types of queries shows that \sys{} can run at more than 100$\times$ video realtime under typical wireless network and camera hardware.

%% file: main.bbl
\begin{thebibliography}{100}

\bibitem{eu-video-guideline}
The european data protection supervisor video-surveillance guidelines.
\newblock
  \url{https://edps.europa.eu/sites/edp/files/publication/10-03-17_video-surveillance_guidelines_en.pdf},
  2010.

\bibitem{surveillance-policy}
Tufts: Video security university policy.
\newblock \url{https://publicsafety.tufts.edu/policies/video-security/}, 2014.

\bibitem{complain-camera-1}
Wireless cameras slowing router too much.
\newblock
  \url{https://community.netgear.com/t5/Nighthawk-WiFi-Routers/Wireless-cameras-slowing-router-too-much/td-p/513047},
  2015.

\bibitem{fortinet-white-paper}
Understanding ip surveillance camera bandwidth.
\newblock
  \url{https://www.fortinet.com/content/dam/fortinet/assets/white-papers/wp-ip-surveillance-camera.pdf},
  2017.

\bibitem{cisco-white-paper}
The zettabyte era: Trends and analysis.
\newblock
  \url{https://www.cisco.com/c/en/us/solutions/collateral/service-provider/visual-networking-index-vni/vni-hyperconnectivity-wp.html},
  2017.

\bibitem{camera-survey}
International trends in video surveillance.
\newblock
  \url{https://cms.uitp.org/wp/wp-content/uploads/2020/06/18-07Statistics-Brief-Videosurveillance-web.pdf},
  2018.

\bibitem{retention-case-law}
New case law on retention periods for video surveillance at the workplace.
\newblock
  \url{https://www.twobirds.com/en/news/articles/2018/germany/new-case-law-on-retention-periods-for-video-surveillance-at-the-workplace},
  2018.

\bibitem{yolov2-rpi}
Running yolo detection on raspberry pi.
\newblock
  \url{http://raspberrypi4u.blogspot.com/2018/10/raspberry-pi-yolo-real-time-object.html},
  2018.

\bibitem{wifi-state}
The state of wifi vs mobile network experience as 5g arrives.
\newblock
  \url{https://www.opensignal.com/sites/opensignal-com/files/data/reports/global/data-2018-11/state_of_wifi_vs_mobile_opensignal_201811.pdf},
  2018.

\bibitem{surveillance-policy2}
Video surveillance laws: Video retention requirements by state.
\newblock
  \url{https://www.verkada.com/blog/surveillance-laws-video-retention-requirements/},
  2018.

\bibitem{complain-camera-2}
Wifi cameras.
\newblock
  \url{https://www.security-camera-warehouse.com/ip-camera/wifi-enabled/},
  2018.

\bibitem{bng-sub}
Background subtraction.
\newblock
  \url{https://docs.opencv.org/3.4.0/db/d5c/tutorial_py_bg_subtraction.html},
  2019.

\bibitem{coral-camera}
Build intelligent ideas with our platform for local ai.
\newblock \url{https://coral.withgoogle.com/}, 2019.

\bibitem{Comcast-dataplan}
Comcast business internet data plan.
\newblock
  \url{https://www.business.org/services/internet/comcast-business-internet-review/},
  2019.

\bibitem{cheapcam}
Hisilicon ip camera specifications.
\newblock \url{http://www.hisilicon.com/en/Products/ProductList/Surveillance},
  2019.

\bibitem{wyzecam}
Wyze camera specifications.
\newblock \url{https://www.wyze.com/wyze-cam/specs/}, 2019.

\bibitem{wyze-camv2}
Wyze camera v2 1080p.
\newblock \url{https://www.wyze.com/product/wyze-cam-v2/}, 2019.

\bibitem{yi-cam}
Yi home camera.
\newblock
  \url{https://www.amazon.com/YI-Security-Surveillance-Monitor-Android/dp/B01CW4AR9K},
  2019.

\bibitem{Ashland}
Youtube live streaming: Ashland.
\newblock \url{https://www.youtube.com/watch?v=e47XhLmZhFk}, 2019.

\bibitem{Banff}
Youtube live streaming: Banff.
\newblock \url{https://youtu.be/9HwSNgcdQ7k}, 2019.

\bibitem{BoatHouse}
Youtube live streaming: Boathouse.
\newblock \url{https://www.youtube.com/watch?v=TXw7CyY0TbU&t=0s}, 2019.

\bibitem{Chaweng}
Youtube live streaming: Chaweng.
\newblock \url{https://www.youtube.com/watch?v=tihJ58_qiH0}, 2019.

\bibitem{CoralReef}
Youtube live streaming: Coralreef.
\newblock \url{https://youtu.be/WYOe8SfQbac}, 2019.

\bibitem{Eagle}
Youtube live streaming: Eagle.
\newblock \url{https://www.youtube.com/watch?v=Q_OrM8o2k6I}, 2019.

\bibitem{JacksonH}
Youtube live streaming: Jackson hole.
\newblock \url{https://youtu.be/2wnU2Kp7quQ}, 2019.

\bibitem{JacksonT}
Youtube live streaming: Jackson town.
\newblock \url{https://www.youtube.com/watch?v=1EiC9bvVGnk}, 2019.

\bibitem{Lausanne}
Youtube live streaming: Lausanne.
\newblock \url{https://www.youtube.com/watch?v=7uF7DsUQ9vc}, 2019.

\bibitem{Miami}
Youtube live streaming: Miami.
\newblock \url{https://www.youtube.com/watch?v=0dctq-YjAdc}, 2019.

\bibitem{Mierlo}
Youtube live streaming: Mierlo.
\newblock \url{https://www.youtube.com/watch?v=HbtBgxFkDHU}, 2019.

\bibitem{Oxford}
Youtube live streaming: Oxford.
\newblock \url{https://www.youtube.com/watch?v=St7aTfoIdYQ}, 2019.

\bibitem{Shibuya}
Youtube live streaming: Shibuya.
\newblock \url{https://youtu.be/PmrWwYTlAVQ}, 2019.

\bibitem{Venice}
Youtube live streaming: Venice.
\newblock \url{https://www.youtube.com/watch?v=JqUREqYduHw}, 2019.

\bibitem{Whitebay}
Youtube live streaming: Whitebay.
\newblock \url{https://www.youtube.com/watch?v=LXWVYoBluT4}, 2019.

\bibitem{zosi-cam}
Zosi camera.
\newblock
  \url{https://www.amazon.com/ZOSI-1280TVL-Security-Weatherproof-Surveillance/dp/B01DF6LJZK},
  2019.

\bibitem{lora}
Alo{\"y}s Augustin, Jiazi Yi, Thomas Clausen, and William Townsley.
\newblock A study of lora: Long range \& low power networks for the internet of
  things.
\newblock {\em Sensors}, 16(9):1466, 2016.

\bibitem{beymer1997real}
David Beymer, Philip McLauchlan, Benjamin Coifman, and Jitendra Malik.
\newblock A real-time computer vision system for measuring traffic parameters.
\newblock In {\em Proceedings of IEEE computer society conference on computer
  vision and pattern recognition (CVPR)}, pages 495--501, 1997.

\bibitem{sparsesampling}
T.~{Blu}, P.~{Dragotti}, M.~{Vetterli}, P.~{Marziliano}, and L.~{Coulot}.
\newblock Sparse sampling of signal innovations.
\newblock {\em IEEE Signal Processing Magazine}, 25(2):31--40, March 2008.

\bibitem{similarity-join}
Christian B{\"o}hm, Bernhard Braunm{\"u}ller, Florian Krebs, and Hans-Peter
  Kriegel.
\newblock Epsilon grid order: An algorithm for the similarity join on massive
  high-dimensional data.
\newblock In {\em ACM SIGMOD Record}, volume~30, pages 379--388, 2001.

\bibitem{pedestrian-cascade}
Zhaowei Cai, Mohammad Saberian, and Nuno Vasconcelos.
\newblock Learning complexity-aware cascades for deep pedestrian detection.
\newblock In {\em Proceedings of the IEEE International Conference on Computer
  Vision (ICCV)}, pages 3361--3369, 2015.

\bibitem{filterforward}
Christopher Canel, Thomas Kim, Giulio Zhou, Conglong Li, Hyeontaek Lim,
  David~G. Andersen, Michael Kaminsky, and Subramanya~R. Dulloor.
\newblock Scaling video analytics on constrained edge nodes.
\newblock In {\em Proceedings of the 2nd SysML Conference (SysML)}, 2019.

\bibitem{queryrefinement}
Kaushik Chakrabarti, Kriengkrai Porkaew, and Sharad Mehrotra.
\newblock Efficient query refinement in multimedia databases.
\newblock In {\em ICDE Conference}, January 2000.
\newblock Poster paper.

\bibitem{glimpse}
Tiffany Yu-Han Chen, Lenin~S. Ravindranath, Shuo Deng, Paramvir~Victor Bahl,
  and Hari Balakrishnan.
\newblock {Glimpse: Continuous, Real-Time Object Recognition on Mobile
  Devices}.
\newblock In {\em 13th ACM Conference on Embedded Networked Sensor Systems
  (SenSys)}, November 2015.

\bibitem{olamr}
Tyson Condie, Neil Conway, Peter Alvaro, Joseph~M. Hellerstein, Khaled
  Elmeleegy, and Russell Sears.
\newblock Mapreduce online.
\newblock In {\em Proceedings of the 7th USENIX Conference on Networked Systems
  Design and Implementation (NSDI)}, pages 21--21, 2010.

\bibitem{secs}
Boyuan Feng, Kun Wan, Shu Yang, and Yufei Ding.
\newblock {SECS:} efficient deep stream processing via class skew dichotomy.
\newblock {\em CoRR}, abs/1809.06691, 2018.

\bibitem{eureka}
Ziqiang Feng, Shilpa George, Jan Harkes, Padmanabhan Pillai, Roberta Klatzky,
  and Mahadev Satyanarayanan.
\newblock Eureka: Edge-based discovery of training data for machine learning.
\newblock {\em IEEE Internet Computing}, PP:1--1, 01 2019.

\bibitem{eva}
Ziqiang Feng, Junjue Wang, Jan Harkes, Padmanabhan Pillai, and Mahadev
  Satyanarayanan.
\newblock Eva: An efficient system for exploratory video analysis.
\newblock {\em SysML}, 2018.

\bibitem{mpdash}
Bo~Han, Feng Qian, Lusheng Ji, and Vijay Gopalakrishnan.
\newblock Mp-dash: Adaptive video streaming over preference-aware multipath.
\newblock In {\em Proceedings of the 12th International on Conference on
  Emerging Networking EXperiments and Technologies (CoNEXT)}, pages 129--143,
  New York, NY, USA, 2016. ACM.

\bibitem{encyclopaediaofmathematics}
Michiel Hazewinkel.
\newblock {\em Encyclopaedia of Mathematics}.
\newblock Springer Netherlands, 1988.

\bibitem{control}
J.~M. {Hellerstein}, R.~{Avnur}, A.~{Chou}, C.~{Hidber}, C.~{Olston},
  V.~{Raman}, T.~{Roth}, and P.~J. {Haas}.
\newblock Interactive data analysis: the control project.
\newblock {\em Computer}, 32(8):51--59, Aug 1999.

\bibitem{ola}
Joseph~M. Hellerstein, Peter~J. Haas, and Helen~J. Wang.
\newblock Online aggregation.
\newblock In {\em Proceedings of the 1997 ACM SIGMOD International Conference
  on Management of Data (SIGMOD)}, pages 171--182, New York, NY, USA, 1997.
  ACM.

\bibitem{focus}
Kevin Hsieh, Ganesh Ananthanarayanan, Peter Bodik, Shivaram Venkataraman,
  Paramvir Bahl, Matthai Philipose, Phillip~B. Gibbons, and Onur Mutlu.
\newblock Focus: Querying large video datasets with low latency and low cost.
\newblock In {\em 13th {USENIX} Symposium on Operating Systems Design and
  Implementation ({OSDI} 18)}, Carlsbad, CA, 2018. {USENIX} Association.

\bibitem{videoedge}
Chien-Chun Hung, Ganesh Ananthanarayanan, Peter Bodík, Leana Golubchik, Minlan
  Yu, Victor Bahl, and Matthai Philipose.
\newblock Videoedge: Processing camera streams using hierarchical clusters.
\newblock In {\em 2018 IEEE/ACM Symposium on Edge Computing (SEC)}, 2018.

\bibitem{rank-aware-opt}
Ihab~F Ilyas, Rahul Shah, Walid~G Aref, Jeffrey~Scott Vitter, and Ahmed~K
  Elmagarmid.
\newblock Rank-aware query optimization.
\newblock In {\em Proceedings of the 2004 ACM SIGMOD international conference
  on Management of data (ICMD)}, pages 203--214, 2004.

\bibitem{rexcam}
Samvit Jain, Junchen Jiang, Yuanchao Shu, Ganesh Ananthanarayanan, and Joseph
  Gonzalez.
\newblock Rexcam: Resource-efficient, cross-camera video analytics at
  enterprise scale.
\newblock {\em CoRR}, abs/1811.01268, 2018.

\bibitem{chameleon}
Junchen Jiang, Ganesh Ananthanarayanan, Peter Bodik, Siddhartha Sen, and Ion
  Stoica.
\newblock Chameleon: Scalable adaptation of video analytics.
\newblock In {\em Proceedings of the 2018 Conference of the ACM Special
  Interest Group on Data Communication (SIGCOMM)}, pages 253--266, New York,
  NY, USA, 2018. ACM.

\bibitem{split-NN}
Tian Jin and Seokin Hong.
\newblock Split-cnn: Splitting window-based operations in convolutional neural
  networks for memory system optimization.
\newblock In {\em Proceedings of the Twenty-Fourth International Conference on
  Architectural Support for Programming Languages and Operating Systems
  (ASPLOS)}, pages 835--847, 2019.

\bibitem{activelearning}
Christoph K{\"a}ding, Erik Rodner, Alexander Freytag, and Joachim Denzler.
\newblock Fine-tuning deep neural networks in continuous learning scenarios.
\newblock In Chu-Song Chen, Jiwen Lu, and Kai-Kuang Ma, editors, {\em Computer
  Vision -- ACCV 2016 Workshops}, pages 588--605, Cham, 2017. Springer
  International Publishing.

\bibitem{blazeit}
Daniel Kang, Peter Bailis, and Matei Zaharia.
\newblock Blazeit: Fast exploratory video queries using neural networks.
\newblock {\em arXiv preprint arXiv:1805.01046}, 2018.

\bibitem{noscope}
Daniel Kang, John Emmons, Firas Abuzaid, Peter Bailis, and Matei Zaharia.
\newblock Noscope: Optimizing neural network queries over video at scale.
\newblock {\em Proc. VLDB Endow.}, 10(11):1586--1597, August 2017.

\bibitem{kang16cvpr}
K.~{Kang}, W.~{Ouyang}, H.~{Li}, and X.~{Wang}.
\newblock Object detection from video tubelets with convolutional neural
  networks.
\newblock In {\em 2016 IEEE Conference on Computer Vision and Pattern
  Recognition (CVPR)}, pages 817--825, June 2016.

\bibitem{koudas2000high}
Nick Koudas and Kenneth~C Sevcik.
\newblock High dimensional similarity joins: Algorithms and performance
  evaluation.
\newblock {\em IEEE Transactions on Knowledge and Data Engineering (TKDE)},
  12(1):3--18, 2000.

\bibitem{deeplens}
Sanjay Krishnan, Adam Dziedzic, and Aaron~J. Elmore.
\newblock Deeplens: Towards a visual data management system.
\newblock In {\em {CIDR} 2019, 9th Biennial Conference on Innovative Data
  Systems Research, Asilomar, CA, USA, January 13-16, 2019, Online
  Proceedings}, 2019.

\bibitem{alexnet}
Alex Krizhevsky, Ilya Sutskever, and Geoffrey~E Hinton.
\newblock Imagenet classification with deep convolutional neural networks.
\newblock In F.~Pereira, C.~J.~C. Burges, L.~Bottou, and K.~Q. Weinberger,
  editors, {\em Advances in Neural Information Processing Systems (NIPS)},
  pages 1097--1105. Curran Associates, Inc., 2012.

\bibitem{cornernet}
Hei Law and Jia Deng.
\newblock Cornernet: Detecting objects as paired keypoints.
\newblock {\em International Journal of Computer Vision (IJCV)}, Aug 2019.

\bibitem{reducto}
Yuanqi Li, Arthi Padmanabhan, Pengzhan Zhao, Yufei Wang, Guoqing~Harry Xu, and
  Ravi Netravali.
\newblock Reducto: On-camera filtering for resource-efficient real-time video
  analytics.
\newblock In {\em Proceedings of the Annual conference of the ACM Special
  Interest Group on Data Communication on the applications, technologies,
  architectures, and protocols for computer communication (SIGCOMM)}, pages
  359--376, 2020.

\bibitem{nnbenchmarks}
Mike Liao.
\newblock Benchmarking hardware for cnn inference in 2018.
\newblock
  \url{https://towardsdatascience.com/benchmarking-hardware-for-cnn-inference-in-2018-1d58268de12a},
  2018.

\bibitem{lipton2015video}
Alan~J Lipton, Peter~L Venetianer, Niels Haering, Paul~C Brewer, Weihong Yin,
  Zhong Zhang, Li~Yu, Yongtong Hu, Gary~W Myers, Andrew~J Chosak, et~al.
\newblock Video analytics for retail business process monitoring, 2015.
\newblock US Patent 9,158,975.

\bibitem{liu17cvpr}
Mason Liu and Menglong Zhu.
\newblock Mobile video object detection with temporally-aware feature maps.
\newblock {\em Proceedings of the IEEE Conference on Computer Vision and
  Pattern Recognition (CVPR)}, 2018.

\bibitem{spectrum-crunch}
{NIST}.
\newblock The spectrum crunch.
\newblock
  \url{https://www.nist.gov/topics/advanced-communications/spectrum-crunch},
  2019.

\bibitem{pakha18hotcloud}
Chrisma Pakha, Aakanksha Chowdhery, and Junchen Jiang.
\newblock Reinventing video streaming for distributed vision analytics.
\newblock In {\em 10th {USENIX} Workshop on Hot Topics in Cloud Computing
  (HotCloud 18)}, Boston, MA, 2018. {USENIX} Association.

\bibitem{online-mapreduce}
Niketan Pansare, Vinayak~R Borkar, Chris Jermaine, and Tyson Condie.
\newblock Online aggregation for large mapreduce jobs.
\newblock {\em Proceedings of the VLDB Endowment}, 4(11):1135--1145, 2011.

\bibitem{wdblog}
Ziv Paz.
\newblock Innovation in surveillance: What’s changing at the edge, core and
  cloud?
\newblock
  \url{https://blog.westerndigital.com/innovation-surveillance-edge-core-cloud/},
  year = {2018}.

\bibitem{scanner}
Alex Poms, Will Crichton, Pat Hanrahan, and Kayvon Fatahalian.
\newblock Scanner: Efficient video analysis at scale.
\newblock {\em ACM Trans. Graph.}, 37(4):138:1--138:13, July 2018.

\bibitem{iridium}
Qifan Pu, Ganesh Ananthanarayanan, Peter Bodik, Srikanth Kandula, Aditya
  Akella, Paramvir Bahl, and Ion Stoica.
\newblock Low latency geo-distributed data analytics.
\newblock {\em SIGCOMM Comput. Commun. Rev.}, 45(4):421--434, August 2015.

\bibitem{jetstream}
Ariel Rabkin, Matvey Arye, Siddhartha Sen, Vivek~S. Pai, and Michael~J.
  Freedman.
\newblock Aggregation and degradation in jetstream: Streaming analytics in the
  wide area.
\newblock In {\em 11th {USENIX} Symposium on Networked Systems Design and
  Implementation ({NSDI} 14)}, pages 275--288, Seattle, WA, 2014. {USENIX}
  Association.

\bibitem{deepdecision}
X.~{Ran}, H.~{Chen}, X.~{Zhu}, Z.~{Liu}, and J.~{Chen}.
\newblock Deepdecision: A mobile deep learning framework for edge video
  analytics.
\newblock In {\em IEEE Conference on Computer Communications (INFOCOM)}, pages
  1421--1429, April 2018.

\bibitem{yolov3}
Joseph Redmon and Ali Farhadi.
\newblock Yolov3: An incremental improvement.
\newblock {\em arXiv preprint arXiv:1804.02767}, 2018.

\bibitem{saligrama12cvpr}
Venkatesh Saligrama and Zhu Chen.
\newblock Video anomaly detection based on local statistical aggregates.
\newblock {\em 2012 IEEE Conference on Computer Vision and Pattern Recognition
  (CVPR)}, pages 2112--2119, 2012.

\bibitem{regularsampling}
Cosma~Rohilla Shalizi.
\newblock Advanced data analysis from an elementary point of view.
\newblock \url{http://www.stat.cmu.edu/~cshalizi/ADAfaEPoV/ADAfaEPoV.pdf}, year
  = {2019}.

\bibitem{shen2019nexus}
Haichen Shen, Lequn Chen, Yuchen Jin, Liangyu Zhao, Bingyu Kong, Matthai
  Philipose, Arvind Krishnamurthy, and Ravi Sundaram.
\newblock Nexus: a gpu cluster engine for accelerating dnn-based video
  analysis.
\newblock In {\em Proceedings of the 27th ACM Symposium on Operating Systems
  Principles (SOSP)}, pages 322--337, 2019.

\bibitem{shen17cvpr}
Haichen Shen, Seungyeop Han, Matthai Philipose, and Arvind Krishnamurthy.
\newblock Fast video classification via adaptive cascading of deep models.
\newblock In {\em The IEEE Conference on Computer Vision and Pattern
  Recognition (CVPR)}, July 2017.

\bibitem{shi2018geometry}
Honghui Shi.
\newblock Geometry-aware traffic flow analysis by detection and tracking.
\newblock In {\em Proceedings of the IEEE Conference on Computer Vision and
  Pattern Recognition (CVPR) Workshops}, pages 116--120, 2018.

\bibitem{clickbait}
Ervin Teng, Jo{\~{a}}o~Diogo Falc{\~{a}}o, and Bob Iannucci.
\newblock Clickbait: Click-based accelerated incremental training of
  convolutional neural networks.
\newblock {\em CoRR}, abs/1709.05021, 2017.

\bibitem{clickbaitv2}
Ervin Teng, Rui Huang, and Bob Iannucci.
\newblock Clickbait-v2: Training an object detector in real-time.
\newblock {\em CoRR}, abs/1803.10358, 2018.

\bibitem{objectdetect-cascade}
Paul Viola, Michael Jones, et~al.
\newblock Rapid object detection using a boosted cascade of simple features.
\newblock {\em Proceedings of the 2001 IEEE computer society conference on
  computer vision and pattern recognition (CVPR)}, 1:511--518, 2001.

\bibitem{clarinet}
Raajay Viswanathan, Ganesh Ananthanarayanan, and Aditya Akella.
\newblock {CLARINET}: Wan-aware optimization for analytics queries.
\newblock In {\em 12th {USENIX} Symposium on Operating Systems Design and
  Implementation ({OSDI} 16)}, pages 435--450, Savannah, GA, 2016. {USENIX}
  Association.

\bibitem{vulimiri15nsdi}
Ashish Vulimiri, Carlo Curino, P.~Brighten Godfrey, Thomas Jungblut, Jitu
  Padhye, and George Varghese.
\newblock Global analytics in the face of bandwidth and regulatory constraints.
\newblock In {\em 12th {USENIX} Symposium on Networked Systems Design and
  Implementation ({NSDI} 15)}, pages 323--336, Oakland, CA, 2015. {USENIX}
  Association.

\bibitem{wanalytics}
Ashish Vulimiri, Carlo Curino, Philip~Brighten Godfrey, Thomas Jungblut,
  Konstantinos Karanasos, Jitendra Padhye, and George Varghese.
\newblock Wanalytics: Geo-distributed analytics for a data intensive world.
\newblock In {\em Proceedings of the 2015 ACM SIGMOD International Conference
  on Management of Data (SIGMOD)}, pages 1087--1092, New York, NY, USA, 2015.
  ACM.

\bibitem{trafficpercentage}
Haizhong Wang, Kimberly Rudy, Jia Li, and Daiheng Ni.
\newblock Calculation of traffic flow breakdown probability to optimize link
  throughput.
\newblock {\em Applied Mathematical Modelling}, 34(11):3376 -- 3389, 2010.

\bibitem{lube}
Hao Wang and Baochun Li.
\newblock Lube: Mitigating bottlenecks in wide area data analytics.
\newblock In {\em 9th {USENIX} Workshop on Hot Topics in Cloud Computing
  (HotCloud 17)}, Santa Clara, CA, 2017. {USENIX} Association.

\bibitem{wang18sec}
Junjue Wang, Ziqiang Feng, Zhuo Chen, Shilpa George, Mihir Bala, Padmanabhan
  Pillai, Shao{-}Wen Yang, and Mahadev Satyanarayanan.
\newblock Bandwidth-efficient live video analytics for drones via edge
  computing.
\newblock In {\em 2018 {IEEE/ACM} Symposium on Edge Computing (SEC)}, pages
  159--173, 2018.

\bibitem{cloudseg}
Yiding Wang, Weiyan Wang, Junxue Zhang, Junchen Jiang, and Kai Chen.
\newblock Bridging the edge-cloud barrier for real-time advanced vision
  analytics.
\newblock In {\em 11th {USENIX} Workshop on Hot Topics in Cloud Computing
  (HotCloud 19)}, 2019.

\bibitem{terrorismpercentage}
Slate William~Saletan.
\newblock The case for mass surveillance.
\newblock
  \url{https://www.delcotimes.com/news/the-case-for-mass-surveillance/article_61a27a3c-8e8b-54e3-b048-e44682b6a024.html},
  2013.

\bibitem{elfcam}
Mengwei Xu, Xiwen Zhang, Yunxin Liu, Gang Huang, Xuanzhe Liu, and Felix~Xiaozhu
  Lin.
\newblock Approximate query service on autonomous iot cameras.
\newblock In {\em Proceedings of the 18th International Conference on Mobile
  Systems, Applications, and Services}, pages 191--205, 2020.

\bibitem{xu2018deepcache}
Mengwei Xu, Mengze Zhu, Yunxin Liu, Felix~Xiaozhu Lin, and Xuanzhe Liu.
\newblock Deepcache: Principled cache for mobile deep vision.
\newblock In {\em Proceedings of the 24th Annual International Conference on
  Mobile Computing and Networking}, pages 129--144, 2018.

\bibitem{vstore}
Tiantu Xu, Luis~Materon Botelho, and Felix~Xiaozhu Lin.
\newblock Vstore: A data store for analytics on large videos.
\newblock In {\em Proceedings of the Fourteenth EuroSys Conference 2019
  (EuroSys)}, pages 16:1--16:17, New York, NY, USA, 2019. ACM.

\bibitem{lavea}
S.~Yi, Z.~Hao, Q.~Zhang, Q.~Zhang, W.~Shi, and Q.~Li.
\newblock Lavea: Latency-aware video analytics on edge computing platform.
\newblock In {\em 2017 IEEE 37th International Conference on Distributed
  Computing Systems (ICDCS)}, pages 2573--2574, June 2017.

\bibitem{awstream}
Ben Zhang, Xin Jin, Sylvia Ratnasamy, John Wawrzynek, and Edward~A. Lee.
\newblock Awstream: Adaptive wide-area streaming analytics.
\newblock In {\em Proceedings of the 2018 Conference of the ACM Special
  Interest Group on Data Communication (SIGCOMM)}, pages 236--252, New York,
  NY, USA, 2018. ACM.

\bibitem{videostorm}
Haoyu Zhang, Ganesh Ananthanarayanan, Peter Bodik, Matthai Philipose, Paramvir
  Bahl, and Michael~J. Freedman.
\newblock Live video analytics at scale with approximation and delay-tolerance.
\newblock In {\em 14th {USENIX} Symposium on Networked Systems Design and
  Implementation ({NSDI} 17)}, pages 377--392, Boston, MA, 2017. {USENIX}
  Association.

\bibitem{vigil}
Tan Zhang, Aakanksha Chowdhery, Paramvir~(Victor) Bahl, Kyle Jamieson, and
  Suman Banerjee.
\newblock The design and implementation of a wireless video surveillance
  system.
\newblock In {\em Proceedings of the 21st Annual International Conference on
  Mobile Computing and Networking (MobiCom)}, pages 426--438, New York, NY,
  USA, 2015. ACM.

\bibitem{zhu2017online}
Hongwei Zhu, Farzin Aghdasi, Greg~M Millar, and Stephen~J Mitchell.
\newblock Online learning method for people detection and counting for retail
  stores, 2017.
\newblock US Patent 9,639,747.

\bibitem{zhu18cvpr}
Xizhou Zhu, Jifeng Dai, Lu~Yuan, and Yichen Wei.
\newblock Towards high performance video object detection.
\newblock In {\em Proceedings of the IEEE Conference on Computer Vision and
  Pattern Recognition (CVPR)}, pages 7210--7218. {IEEE} Computer Society, 2018.

\end{thebibliography}
